\documentclass[10pt,journal]{IEEEtran}
\usepackage{array,multirow,graphicx}
\usepackage{graphicx}
\usepackage{epsf}
\usepackage{epsfig}
\usepackage{psfrag}
\usepackage{ifthen}
\usepackage{law}
\usepackage{mystyle}
\usepackage{graphics}
\usepackage{times}
\usepackage{cite}
\usepackage{amsbsy}
\usepackage{amsfonts}
\usepackage{amsmath}
\usepackage{amssymb}
\usepackage{multicol}
\usepackage{float}
\usepackage{steinmetz}
\usepackage{algorithm}
\usepackage{algorithmic}
\usepackage{lettrine}
\usepackage[misc]{ifsym}
\usepackage{color}
\usepackage{todonotes}
\usepackage{caption}
\usepackage{subcaption}

\begin{document}

\title{\Huge Collaborative Pipeline Using Opportunistic Mobile Resources via D2D for Computation-Intensive Tasks}

\author{Terry N. Guo,
Hawzhin Mohammed,
and Syed R. Hasan
\thanks{
\textit{(Corresponding author: Terry N. Guo.)}}
\thanks{T. Guo is with Center for Manufacturing Research, Tennessee Technological University (TTU), Cookeville, TN 38505, USA (Email: nguo@tntech.edu). H. Mohammed and S. Hasan are with Department of Electrical and Computer Engineering, TTU (Email: hmohammed42@students.tntech.edu, shasan@tntech.edu).}
}

\maketitle

\begin{abstract}
This paper proposes a mobile pipeline computing concept in a Device-to-Device (D2D) communication setup and studies related issues, where D2D is likely based on millimeter wave (mmWave) in the 5G mobile communication. The proposed opportunistic system employs a cluster of pipelined resource-limited devices on the move to handle real-time on-site computation-intensive tasks for which current cloud computing technology may not be suitable. The feasibility of such a system can be anticipated as high-speed and low-latency wireless technologies get mature. We present a system model by defining the architecture, basic functions, processes at both system level and pipeline device level. A pipeline path finding algorithm along with a multi-task optimization framework is developed. To minimize the search space since the algorithm may need to be run on resource-limited mobile devices, an adjacency-matrix-power-based graph trimming technique is proposed and validated using simulation. Preliminary feasibility assessment of our proposed techniques is performed using experiment and computer simulation. As part of feasibility assessment, the impact of mmWave blockage on the pipeline stability is analyzed and examined for both single-pipeline and concurrent-multiple-pipeline scenarios. Our design and analysis results provide certain insight to guide system design and lay a foundation for further work in this line.
\end{abstract}


\IEEEpeerreviewmaketitle

\vspace{3mm}
\section{Introduction}
\IEEEPARstart{D}{evice}-to-Device (D2D) communication
\cite{lin2000multihop,fodor2012design,pyattaev2014network,qiao2015enabling,wu2016enabling,lien20163gpp,alim2017leveraging,orsino2017exploiting,giatsoglou2017d2d,ansari20175g,haus2017security,sim20175g,he2019d2d} is a promising technology that allows devices to communicate each other directly, without traversing the core network infrastructures. Spurred by the emerging applications of the 
5G mobile communication \cite{boccardi2013five,rappaport2013millimeter,aazam2019fog}
and millimeter-wave (mmWave) communication \cite{cotton2009millimeter,rappaport2013millimeter,maccartney2017rapid},
D2D is getting renewed attention as it can be well integrated with 5G and mmWave  \cite{qiao2015enabling,giatsoglou2017d2d,ansari20175g,sim20175g}
to increase overall spectral efficiency and potentially improve network throughput, energy efficiency, delay, and fairness. Due to its physical proximity capability, D2D can not only benefit mobile users by extending the cellular coverage, but also enable sharing data and even computing resources among geographically nearby users in a real-time manner. Actually, its feature of low communication latency can enable new avenues such as mobile opportunistic pipeline computing which is proposed and studied in this paper. Pipeline concept is widely used for efficiently processing streaming data with multiple computing engines \cite{duncan1990survey,pautasso2006parallel,lane2016deepx,liu2018demand}.
A typical pipeline computing setup involves a group of wire-connected devices that can be inside a 
chip or on one or multiple circuit boards in a lab room. It is possible to use a cluster of pipelined resource-limited mobile devices for timely computation-intensive tasks like real-time image classification on the move, provided the wireless links between devices meet bandwidth and latency requirements. As mobile computing is gaining popularity, especially as 5G and mmWave become a reality, such mobile pipeline computing can be very promising. The availability of high-speed and low-latency wireless connections opens a door for a higher level of resource sharing applications, since it allows us to share scattered computing resources for bandwidth-hungry computation-intensive applications.

Indeed, extending pipeline computing to mobile environments enables many new applications that are difficult to support by current cloud computing technology. Participated individuals equipped with mobile computing resources can be swarms of drones or collaborative industrial robots \cite{andreev2019dense}. They can also be police squads, first responder teams or network-connected soldiers in a battle field \cite{abdelzaher2018will,cotton2009millimeter}.
Several modern applications require continuous and real time data analysis on high volume of data at a remote location or in a harsh environment or both. Currently, node-level edge devices (NEDs), such as a UAV with sensors and wireless communication module, are used mainly to perform data collection or other straightforward jobs such as extinguishing fires and recording videos \cite{giyenko2016intelligent,motlagh2017uav,kim2018collision,ding2018amateur}. Embedding any intelligence in such NEDs requires the aid of cloud or edge servers, which incurs latency. Usage of cloud and edge servers becomes impractical in such situations, because the remote locations and harsh environments make the communication infrastructure requiring base stations either unavailable or very unstable. With no appropriate communication mechanism in such environments, existing edge intelligence solutions cannot be deployed. Moreover, Internet of battlefield things (IoBT) \cite{abdelzaher2018will} is a new paradigm for Internet of Things (IoT) which also requires intelligent real-time data analysis in difficult-to-access places. Existing processing capabilities of NEDs are too limited to achieve real-time and dynamic Artificial intelligence (AI) inference \cite{mao2017modnn,mao2017mednn,zhao2018deepthings,pacheco2018smart}. To gain from the full potential of AI, it needs to be more readily available to the users. A mobile pipeline computing platform allows the NEDs to enjoy the benefits of AI without incurring the latency and bandwidth limitations for to-and-fro data traffic between the server and the devices.

Inspired by these technology trends and potential applications, 
we propose a mobile opportunistic pipeline computing system that takes advantage of high-speed and low-latency wireless connections
to make use of spare computing resources in a community pool. The system includes three types of elements: 1) \textit{workers} who have spare computing resources and are willing to offer computing services, 2) \textit{job requester} who asks the community to help to complete a computation job, and 3) \textit{system manager} who is responsible for maintaining the system. The proposed system largely differs from regular non-real-time resource sharing and crowdsourcing in the sense that mobile pipeline computing needs a group of connected computing engines working simultaneously in a timely coordinated fashion. The challenges we face come from many aspects: high-speed and low-latency requirements, dynamic characteristics of wireless channels, and uncertainty of mobile resources in terms of resource availability and mobility impact on link connectivity, etc. In this paper we try to address some of these challenges from the following angles:
\begin{itemize}
	\item \textit{\textbf{System architecture}--}Provide a layout of the whole system in a D2D setup.
	\item \textit{\textbf{
	Dynamic computing pool of participants}--}Define a framework for handling a pool of participants.
	\item \textit{\textbf{Operation process and protocol}--}Explain operation details at both system level and mobile pipeline device level.
	\item \textit{\textbf{Efficient pipeline path finding}--}Design and implementation of an optimal yet efficient technique for path finding. 
	\item \textit{\textbf{Impact of mmWave blockage}} \cite{collonge2004influence,
singh2009blockage,jung2016connectivity,maccartney2017rapid}
--Analyze mmWave blockage (a severe cause to poor link quality; mmWave is a good fit for D2D in 5G) impact on pipeline stability.
	\item \textit{\textbf{Feasibility of mobile pipeline computing}--}Build a testbed and obtain firsthand assessment results experimentally.
\end{itemize}

Our major contributions in this paper include:
\begin{enumerate}
	\item Proposed a mobile opportunistic pipeline computing concept that takes advantage of high-speed and low-latency D2D connections.
	\item Designed a pipeline path finding method based on a multi-task optimization framework.
	\item Developed an adjacency-matrix-power-based graph trimming technique for reducing path finding complexity, without significantly affecting performance.
	\item Analyzed the pipeline stability performance based on a dynamic mmWave blockage model, and obtained results for both single-pipeline and concurrent-multiple-pipeline scenarios, which gives some insight to guide system design.
	\item Performed feasibility assessment and performance evaluation using experiment and computer simulation.
\end{enumerate}

The rest of this paper is arranged as follows. The system model under consideration is described in the next section, including the architecture, basic functions, processes at both system level and pipeline device level. Section III systematically introduces the pipeline forming techniques with illustrative explanation. Preliminary feasibility assessment based on experimental and computer simulation is provided in Section IV. Analytical and numerical results of mmWave blockage impact on pipeline stability are presented in Section V. Finally, remarks and future work are summarized in Section VI.

\vspace{3mm}
\section{System Model}
In this section, we provide system model, key functions for system management and operation, system-level process and pipeline control at the device level.

\vspace{3mm}
\subsection{System Architecture}
\begin{figure}[h]
\vspace{5mm}
\setlength{\abovecaptionskip}{1mm}
\setlength{\belowcaptionskip}{1mm}
	\centering
	\includegraphics[width=0.43\textwidth,trim=4 4 4 4,clip]{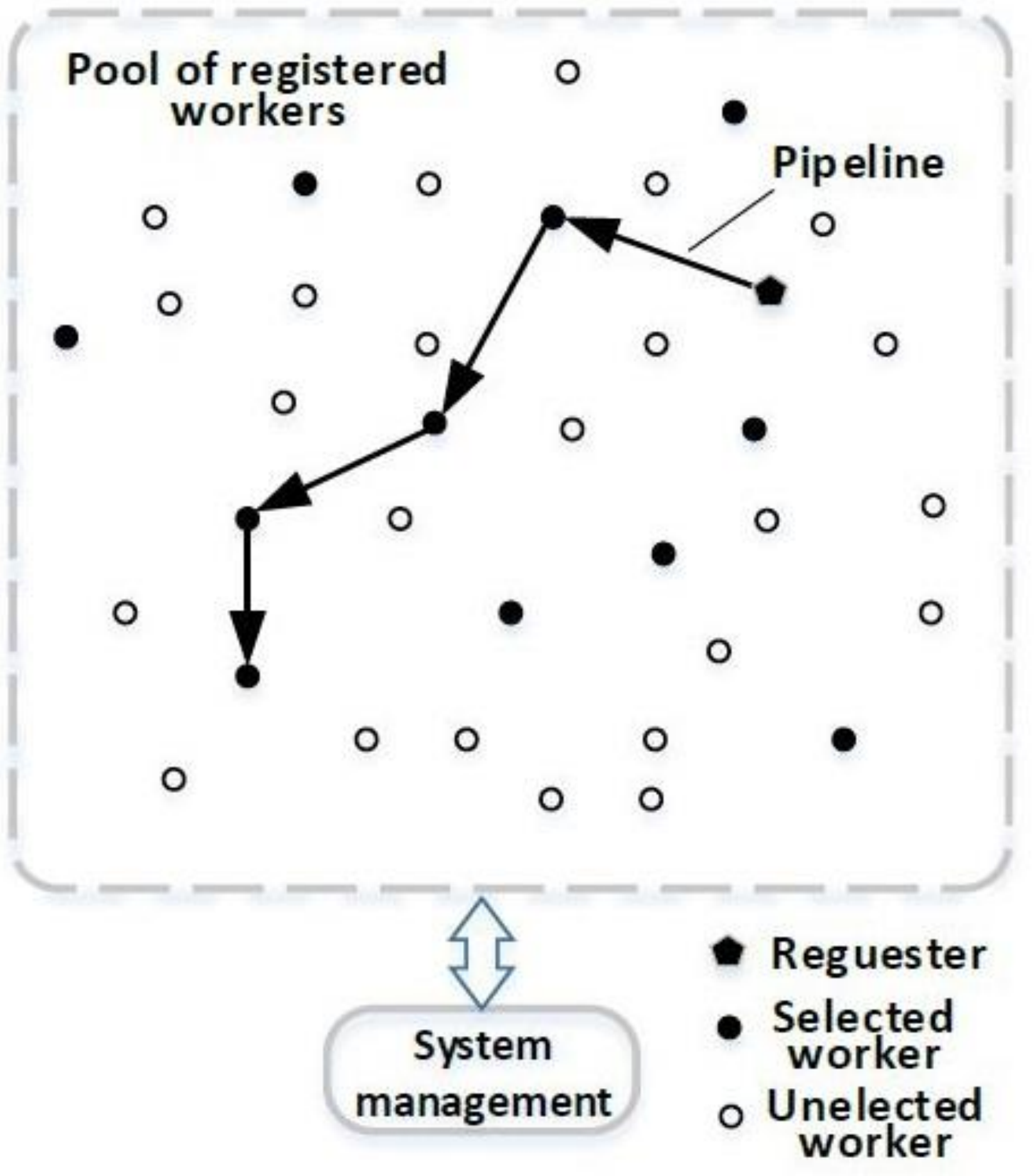}
	\caption{Mobile pipeline computing system.}
\label{fig:system}
\end{figure}

Our proposed mobile pipeline computing system contains three function blocks that are interconnected wirelessly. As illustrated in Fig. \ref{fig:system}, these function blocks are job requesting, system management and worker pooling. A prerequisite for mobile pipeline computing is the existence of a pool of enough registered workers and availability of at least one chain of wideband-connected workers led by the requester. The whole system contains multiple entities, and they play three types of basic roles: job requester, system manager and workers. 
One entity may play multiple roles in reality. However, for the sake of explanation, we assume the job requester does not execute computation tasks as a worker does.

\begin{figure}[h]
\vspace{3mm}
\setlength{\abovecaptionskip}{1mm}
\setlength{\belowcaptionskip}{1mm}
	\centering
	\includegraphics[width=0.5\textwidth,trim=4 4 4 4,clip]{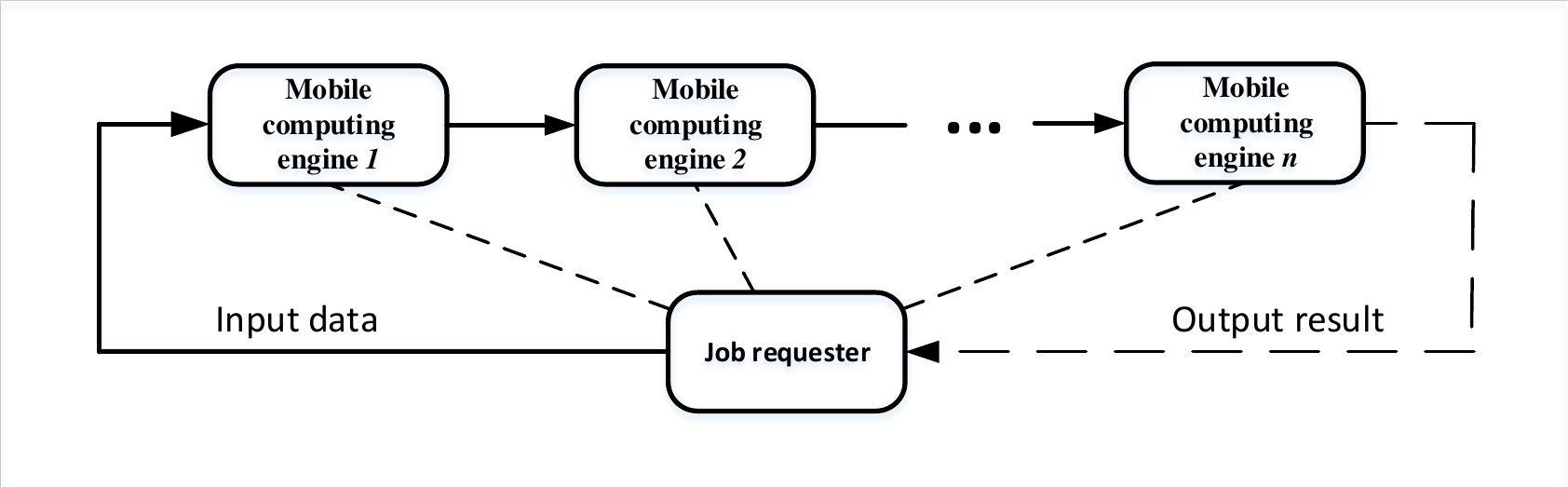}
	\caption{Mobile pipeline computing at device level.}
\label{fig:MobilePipelineComputing}
\end{figure}

At the pipeline device level, the setup of proposed mobile pipeline computing is shown in Fig. \ref{fig:MobilePipelineComputing}. The pipeline is a chain of mobile computing devices connected using wideband communication links.

\subsection{Basic Functions}
The workers, job requester and system manager perform particular functions which are described below.

Workers:
\begin{itemize}
	\item Respond to the system manager and requester
	\item Test and report link qualities
	\item Execute computation jobs
	\item Report job processing status
\end{itemize}

Requester:
\begin{itemize}
	\item Pipeline path search
	\item Data input
	\item Pipeline monitoring
	\item Gather output from the last worker on the pipeline
	\item Report workers' performance to the system manager
\end{itemize}

System manager:
\begin{itemize}
	\item Maintain worker database
	\item Perform worker and requester verification
	\item Calculate and update workers' reliabilities
	\item Recommend qualified workers/nodes
	\item Handle cost and reward
\end{itemize}

There can be different ways to perform system management and maintenance functions. A dedicated resource at an edge server may be reserved for this purpose. It is also possible that these functions are played by a requester, a worker or multiple entities together.
A sustainable opportunistic computing system also requires a reasonable incentive mechanism. These topics are beyond the scope of this paper, and in the following we assume a pool of registered workers have been ready.

\subsection{System Process of Mobile Pipeline Computing}
\begin{figure}[h]
\setlength{\abovecaptionskip}{3mm}
\setlength{\belowcaptionskip}{0mm}
	\centering
	\includegraphics[width=0.5\textwidth]{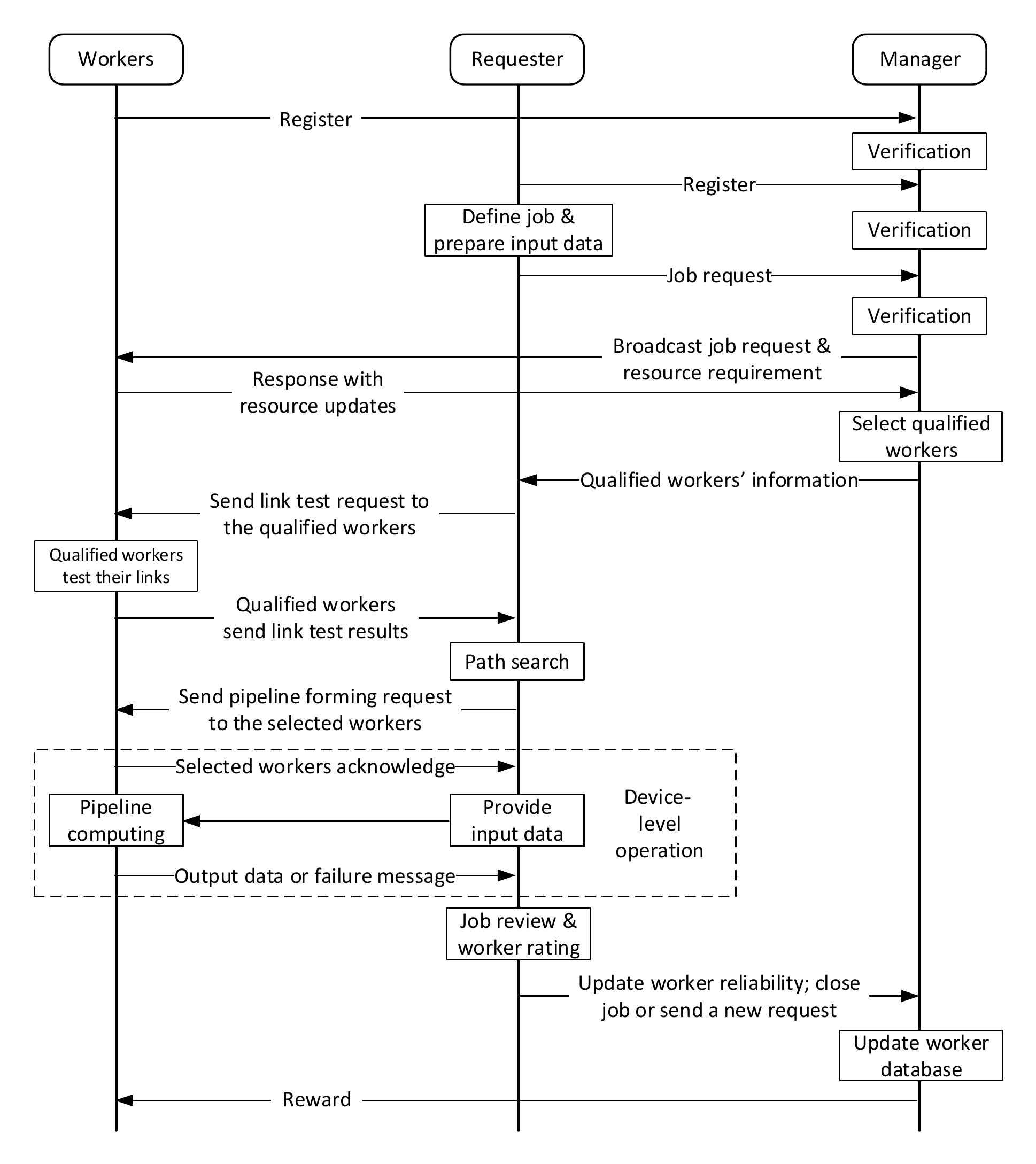}
	\caption{System-level process.}
\label{fig:process}
\end{figure}

The overall system-level process is shown in Fig. \ref{fig:process}. Process at system level covers registration, coordination, and maintenance, etc. To help readers capture the main work flow easily, exceptional events, such as not receiving an expected response, are not shown in the figure. In general, these exceptional events can be taken care of by forcing a respected system element to go to a defaulted state after a time-out limit is reached. The details regrading pipeline operation at device level during a pipeline session (the dash-line box in Fig. \ref{fig:process}) are discussed in next subsection.

\subsection{Pipeline Device-Level Operation--State Machines}

\begin{figure}
\vspace{0mm}
\setlength{\abovecaptionskip}{3mm}
\setlength{\belowcaptionskip}{3mm}
    \centering
\begin{minipage}{0.51\textwidth}
    \centering
	\includegraphics[width=1\textwidth,trim=0.5cm 2.5cm 4.55cm 4,clip]{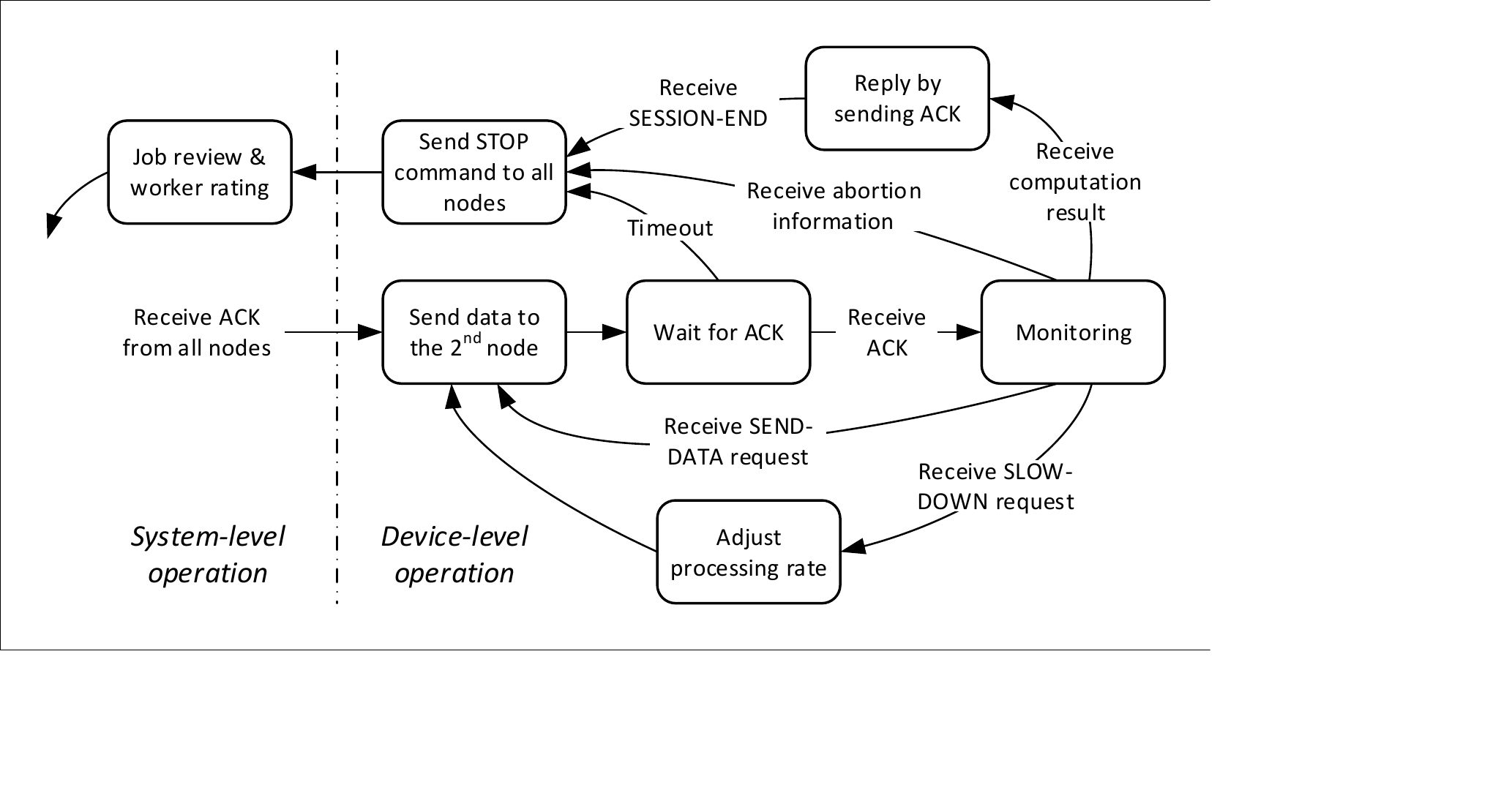}
	\subcaption{Requester state machine.}
\end{minipage}
\begin{minipage}{0.49\textwidth}
    \centering
	\includegraphics[width=1\textwidth]{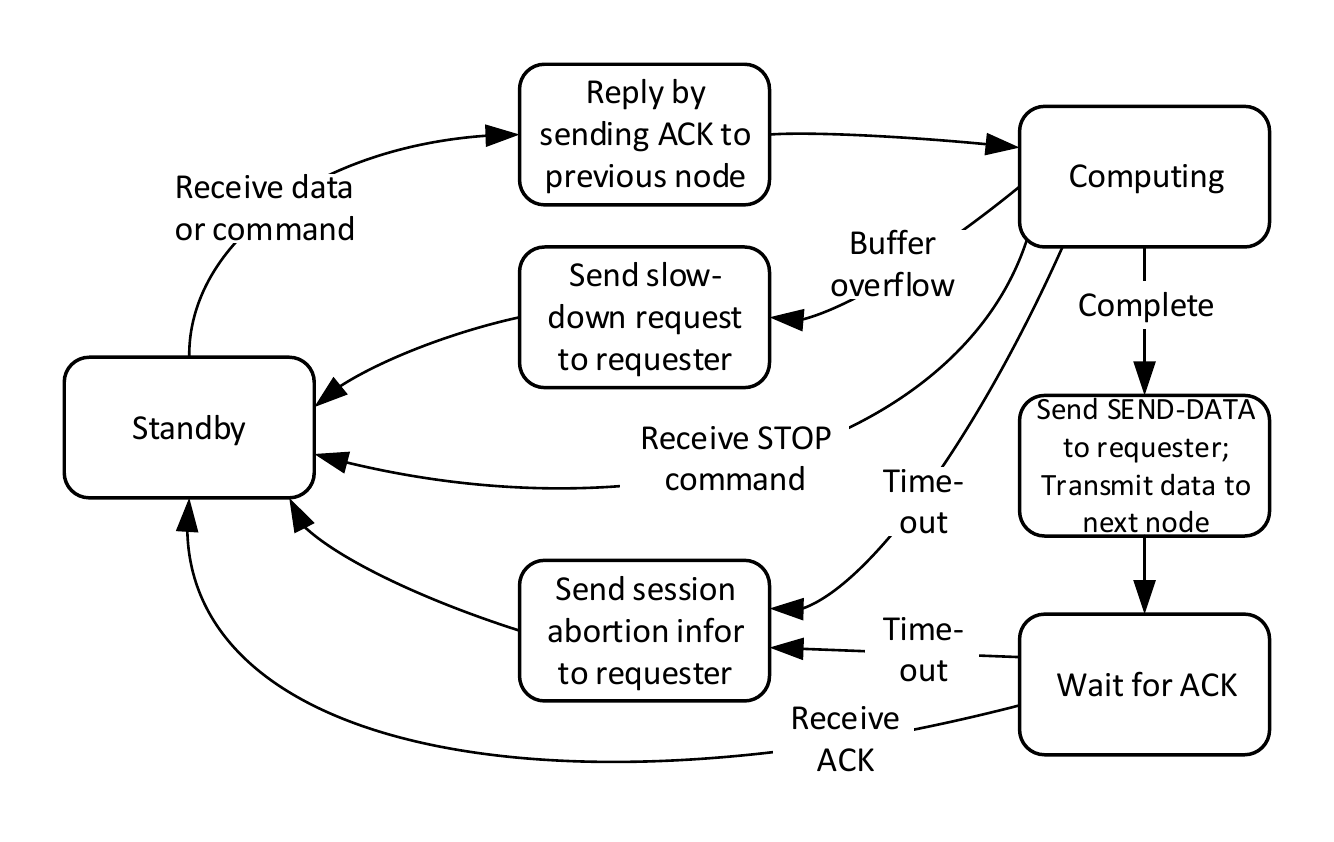}
	\subcaption{Worker state machine.}
\end{minipage}
	\caption{State machines during pipeline session}
\label{fig:state_machine_ACK}
\end{figure}

We describe the device-level operation (the dash-line box in Fig. \ref{fig:process}) by using state machines. Fig. \ref{fig:state_machine_ACK} illustrates the state machines during a pipeline session for the requrester (Fig. \ref{fig:state_machine_ACK} (a)) and each worker (Fig. \ref{fig:state_machine_ACK} (b)). Upon receiving acknowledgments (ACKs) from all workers selected during the previous process, the pipeline is formed and the pipeline session starts. 
At the beginning of a pipeline session, raw data packages are available at the requester and job partitioning is defined. The interactions between the requester and workers can be seen in both Fig. \ref{fig:MobilePipelineComputing} and \ref{fig:state_machine_ACK}. The requester node in a pipeline session is responsible for feeding the data packages to the first worker node, adjusting processing rate upon receiving a request from anyone on the pipeline, and collecting computation outputs from the last worker node on the pipeline. On the other hand, each worker node is responsible for accurately receiving data packages from the node in the front of it, executing its part of computation, and pass the computation outputs to the next node. In addition, when buffer overflowing or timeout happen at any worker node, the node informs the requester node to take actions accordingly.

\section{Mobile Collaborative Pipeline Forming}
Reliable communication links are the essence for mobile pipeline computing. These links can be categorized into two types of channels: wideband forward channels (solid lines in Fig. \ref{fig:MobilePipelineComputing}) for transferring data from one to another, and narrowband channels for monitoring, controlling and sending output data. It is reasonable to assume that the narrowband channels rely on regular communication infrastructures and are available at a high probability. For pipeline path finding we only consider the wideband forward channels, where path finding is an optimization process to find out the best path associated with a given number of workers.
For simplicity we use a worker's reliability to jointly represent his/her job service quality and trustworthiness level. A trustworthiness score can be measured quantitatively and updated sequentially with a forgetting factor or learning rate \cite{wang2003trust,chen2016trust,wang2018trust}.
The concept used in \cite{wang2003trust,chen2016trust,wang2018trust} may be adopted here to calculate the reliability value of each participating worker. We assume the reliability score of each worker is available prior to path finding process.

\subsection{Pipeline Path Finding--Optimization Framework}

The requester has a strategy table that specifies minimum resource requirement for each pipeline configuration. Specifically, a strategy is a set of minimum required computation and communication resources, thus it can be used as a criteria for eliminating some unqualified nodes and links. Table \ref{tab:strategies} shows notations of all  $I$ strategies. Strategy $i\in\{1, 2, \cdots, I\}$ has a predefined preference score $f_i$ and needs $n_i$ distributed computing workers along with a job requester (treated as a special worker with an index label $(i,0)$) to form a pipeline. For strategy $i\in\{1, 2, \cdots, I\}$, from the first worker to the last worker on the pipeline, they are labeled as $(i,0), (i,1), \cdots$, and $(i,n_i)$, respectively; the individual resource requirements in terms of computing power, memory, buffer and communication bandwidth for the $(i,n)$-th worker are specified by resource indexes $c_{i,n}^{req}$, $m_{i,n}^{req}$, $b_{i,n}^{req}$ and $w_{i,n}^{req}$, respectively. All the resource indexes are normalized values between zero and one. This strategy table can be written in a compact format $\{f_i, n_i, \boldsymbol{c}_i^{req}, \boldsymbol{m}_i^{req}, \boldsymbol{b}_i^{req}, \boldsymbol{w}_i^{req}\}_{i=0}^I$, where $\boldsymbol{c}_i^{req}=(c_{i,0}^{req}, c_{i,1}^{req}, \cdots, c_{i,n_i}^{req})$, $\boldsymbol{m}_i^{req}=(m_{i,0}^{req}, m_{i,1}^{req}, \cdots, m_{i,n_i}^{req})$, $\boldsymbol{b}_i^{req}=(b_{i,0}^{req}, b_{i,1}^{req}, \cdots, b_{i,n_i}^{req})$, and $\boldsymbol{w}_i^{req}=(w_{i,0}^{req}, w_{i,1}^{req}, \cdots, w_{i,n_i}^{req})$ are resource index vectors.

\begin{table}[t]
\caption{Strategy table for Pipeline Path Finding.}
\begin{center}
\begin{tabular}{|c|c|c|c|c|c|}
\hline
	{\tiny\bf Strategy} & {\tiny\bf \# of} & {\tiny\bf Computing} & {\tiny\bf Memory} & {\tiny\bf Buffer} & {\tiny\bf Bandwidth}\\
	{\tiny\bf score} & {\tiny\bf workers} & {\tiny\bf index} & {\tiny\bf index} & {\tiny\bf index} & {\tiny\bf index} \\
\hline
	\multirow{4}{*}{$f_1$} & \multirow{4}{*}{$n_1$} & $c_{1,0}^{req}$ & $m_{1,0}^{req}$ & $b_{1,0}^{req}$ & $w_{1,0}^{req}$ \\
\cline{3-6}
	&  & $c_{1,1}^{req}$ & $m_{1,1}^{req}$ & $b_{1,1}^{req}$ & $w_{1,1}^{req}$ \\
\cline{3-6}
	& & \vdots & \vdots & \vdots & \vdots \\
\cline{3-6}
	&  & $c_{1,n_1}^{req}$ & $m_{1,n_1}^{req}$ & $b_{1,n_1}^{req}$ & $w_{1,n_1}^{req}$ \\
\hline
	\multirow{4}{*}{$f_2$} & \multirow{4}{*}{$n_2$} & $c_{2,0}^{req}$ & $m_{2,0}^{req}$ & $b_{2,0}^{req}$ & $w_{2,0}^{req}$ \\
\cline{3-6}
	&  & $c_{2,1}^{req}$ & $m_{2,1}^{req}$ & $b_{2,1}^{req}$ & $w_{2,1}^{req}$ \\
\cline{3-6}
	& & \vdots & \vdots & \vdots & \vdots \\
\cline{3-6}
	&  & $c_{2,n_2}^{req}$ & $m_{2,n_2}^{req}$ & $b_{2,n_2}^{req}$ & $w_{2,n_2}^{req}$ \\
\hline
	\vdots & \vdots & \vdots & \vdots & \vdots & \vdots \\
\hline
	\multirow{4}{*}{$f_I$} & \multirow{4}{*}{$n_I$} & $c_{I,0}^{req}$ & $m_{I,0}^{req}$ &  $b_{I,0}^{req}$ & $w_{I,0}^{req}$ \\
\cline{3-6}
	&  & $c_{I,1}^{req}$ & $m_{I,1}^{req}$ & $b_{I,1}^{req}$ & $w_{I,1}^{req}$ \\
\cline{3-6}
	& & \vdots & \vdots & \vdots & \vdots \\
\cline{3-6}
	&  & $c_{I,n_I}^{req}$ & $m_{I,n_I}^{req}$ & $b_{I,n_I}^{req}$ & $w_{I,n_I}^{req}$ \\
\hline
\end{tabular}
\end{center}
\label{tab:strategies}
\end{table}

All participating workers form a weighted undirected graph $G_0({\cal N}_0, {\cal E}_0)$ with node (worker) set ${\cal N}_0$ and edge (link) set ${\cal E}_0$, where ``undirected'' refers to symmetric or reciprocal communication links. In the language of graph, workers and links are called nodes and edges, respectively; and ``link'' and ``edge'' are used interchangeably throughout the rest of the paper. A pipeline corresponds to a path on the graph. For each strategy there is a set of qualified paths. We propose a multi-task optimization framework to find the best pair of strategy and associated path.

For edge $(s,t)$, let $q_{s,t}\in(0,1]$ be link connection quality, and $r_k\in [0,1]$ be the reliability score for node $k$. We define link reliability $r_{s,t}$ as a geometric average of the two associated end nodes' reliabilities
\begin{eqnarray}
	&& r_{s,t} = \sqrt{r_s \cdot r_t},\; r_s,r_t\in [0,1]
\end{eqnarray}
where $r_s$ and $r_t$ are the two end nodes' reliabilities, $s\neq t,\;s,t=1,2,\cdots,|{\cal N}_0|$. Link quality and link reliability can be extended to path quality and path reliability, respectively. As what has been used in analyzing the reliability of a series system with independent components \cite{kjerengtroen1984structural}, we define path quality as a product of link qualities of all links on the path; similarly, path reliability is a product of link reliabilities of all links on the path. Label a path by $j$, and let ${\cal P}_i$ be the path set for strategy $i$,  and $Q_j$ and $R_j$ ($j\in{\cal P}_i,\;i=1, 2, \cdots, I$) be the path quality and path reliability of path $j$, respectively. Now the quality of a path can be measured using a path score $\left(\beta_1 Q_j + \beta_2 R_j + \beta_3 f_i \right)$ with $\beta_1,\; \beta_2,\;\beta_3$ being some predefined constants.

In order to identify and remove unqualified edges and nodes in advance, we define minimum link quality requirement $q_{min}$, minimum node reliability requirement $r_{min}$, and minimum resource requirements $\{c_{min}, m_{min}, b_{min}, w_{min}\}$. These minimum requirements serve as thresholds in selection of qualified edges and nodes. After removing unqualified edges and nodes (and possibly performing graph trimming as well), Graph $G_0({\cal N}_0, {\cal E}_0)$ is reduced to a smaller graph $G({\cal N}, {\cal E})$ with node set ${\cal N}$ and edge set ${\cal E}$.
Given $I$ qualified path sets ${\cal P}_i,\;i=1, 2, \cdots, I$, search of the best path can be formulated as a multi-task optimization problem:
\begin{eqnarray}
	&&\hspace{-10mm}\max_{i, j\in{\cal P}_i} \left\{\beta_1 Q_j + \beta_2 R_j + \beta_3 f_i \right\}, \; \beta_1, \beta_2,\beta_3 > 0 \nonumber\\
	&&\hspace{-9mm} \text{subject to: }\{\boldsymbol{c}_i^{req}, \boldsymbol{m}_i^{req}, \boldsymbol{b}_i^{req}, \boldsymbol{w}_i^{req}\} \text{ are satisfied} \\
	&&\hspace{18mm} 1\le i\le I \nonumber
\label{opt}
\end{eqnarray}

\begin{algorithm}[!tp]
\caption{Pipeline path finding.}
\label{alg1}
\begin{algorithmic}
\REQUIRE A weighted undirected graph 
$G({\cal N}, {\cal E})$ with $M$ nodes and a strategy table $\{f_i, n_i, \boldsymbol{c}_i^{req}, \boldsymbol{m}_i^{req}, \boldsymbol{b}_i^{req}, \boldsymbol{w}_i^{req}\}_{i=1}^I$ with $I$ strategies for a given requester. \\
\ENSURE Optimal pipeline path along with its associated pipeline configuration strategy.\\
\hspace{-2mm}\textbf{Initialization} \\
Designate the requester node as the root of a path tree.\\
\FOR {$i=1,2,\cdots, I$}
\IF {the requester node meet resource requirement $\{c_{i,0}^{req}, m_{i,0}^{req}, b_{i,0}^{req}, w_{i,0}^{req}\}$,}
\item Perform \textbf{\textit{Forward Search}} (up-down):
\FOR {$n=1,2,\cdots, n_i$}
\item Find ${\cal B}_n$, a subset of all edges that represent tree branches at search depth $n$.\\
\ENDFOR
\IF {${\cal B}_{n_i}$ is not empty,}
\item Perform \textbf{\textit{Backward Tracing}} (bottom-up)): \\
\item Give all the edges in ${\cal B}_{n_i}$ secondary labels $1,2,\cdots, J_i$, i.e., ${\cal B}_{n_i}=\{e_1, e_2, \cdots, e_J\}$.
\FOR {$j=1,2,\cdots, J_i$}
\item Initialize stack $Path_j$ with $e_j$: $Path_j = [e_j]$ .
\FOR {$n=n_i-1, n_i-2,\cdots, 2$}
\item Grow stack $Path_j$ by taking an edge from ${\cal B}_{n}$ and appending it to $Path_j$ such that the last two edges in $Path_j$ are connected on graph $G({\cal N}, {\cal E})$.
\ENDFOR
\item Delete $Path_j$ if $\{\boldsymbol{c}_i^{req}, \boldsymbol{m}_i^{req}, \boldsymbol{b}_i^{req}, \boldsymbol{w}_i^{req}\}$ is not met ($Path_j$ represents a potential path);\\
\item Calculate and store path score of $Path_j$.
\ENDFOR
\item Find the path with the highest score for strategy $i$ among $Path_j,\;j=1,2,\cdots,J_i$, and label it as path-$i$.
\ENDIF
\ENDIF
\ENDFOR\\
\item Among path-$i,\;i=1,2,\cdots,I$, choose the path and its associated strategy with the highest score. \\
\end{algorithmic}
\hspace{-2mm}\textbf{* Note}: $J_i$ is the maximum number of leaves of the search tree for strategy $i$.
\end{algorithm}

\begin{figure}[t]
\vspace{0mm}
\setlength{\abovecaptionskip}{3mm}
\setlength{\belowcaptionskip}{3mm}
    \centering
\begin{minipage}{0.5\textwidth}
    \centering
	\includegraphics[width=0.75\textwidth,trim=4 4 4 4,clip]{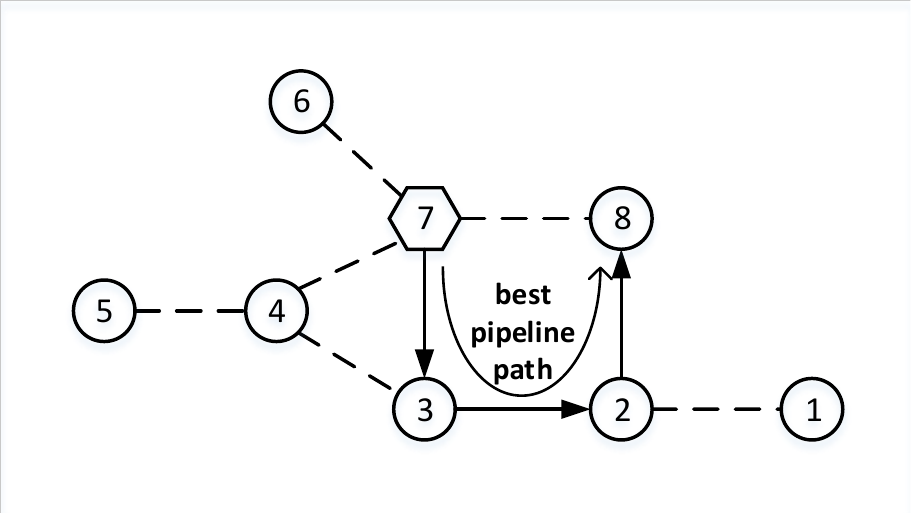}\\
\vspace{0mm}
	\subcaption{A computing cohort with a requester (node 7) and seven potential workers.}
\end{minipage}
\begin{minipage}{0.5\textwidth}
    \centering
	\includegraphics[width=1\textwidth,trim=4 4 4 4,clip]{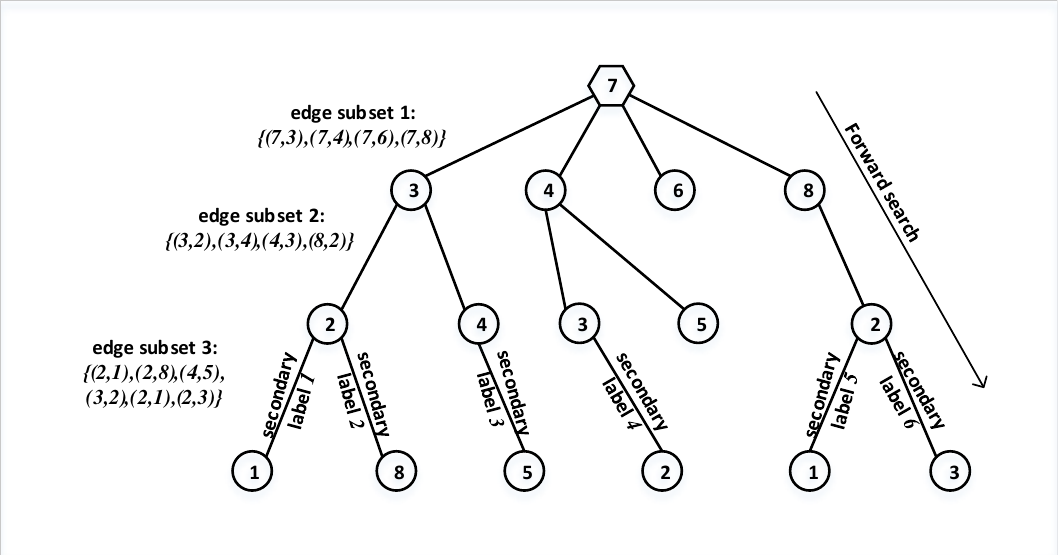}\\
\vspace{0mm}
	\subcaption{\textit{\textbf{Forward Search}} finds all edge subsets on a tree with node 7 as its root.}
\end{minipage}
\begin{minipage}{0.5\textwidth}
\vspace{0mm}
    \centering
	\includegraphics[width=0.95\textwidth]{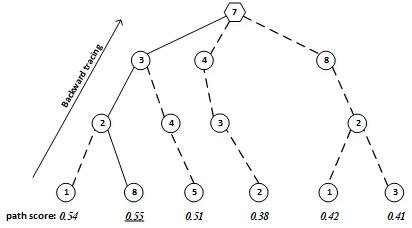}\\
	\subcaption{\textit{\textbf{Backward Tracing}} connects the found edges and finds all potential paths; the winning path for a given strategy has the highest score.}
\end{minipage}
	\caption{A path finding example to find a length-3 ($n_i=3$) pipeline path for a given strategy.}
\label{fig:path_find}
\vspace{0mm}
\end{figure}

\subsection{Pipeline Path Finding Algorithm}
\textbf{Explanation}: Depending on the reliabilities of nodes and links as well as which configuration strategy is chosen, a cohort \footnote{We use ``cohort'' to differentiate it from ``pool'' that contains all registered workers} of selected workers are fed into the path finding process. The path finding process is implemented in an algorithm described in \textbf{Algorithm \ref{alg1}}. To help understand the algorithm, let us use an example to explain it. Fig. \ref{fig:path_find} (a) shows a graph representing a cohort of seven qualified workers including the requester (node 7). Some preprocess may have been done to reduce the graph by eliminating unqualified nodes and edges and performing graph trimming (to be discussed in next subsection). Given the requester node, all feasible pipeline paths actually form a search tree with the requester node as its root (refer to Fig. \ref{fig:path_find} (b) ), where we do not care the portion beyond depth $n_i$ (the required number of workers). Note that we use the tree to help explain, but knowing the tree in advance is not a prerequisite for the algorithm to run. In general, a strategy affects the cohort, thus it affects the search tree as well. Given both the requester node and strategy, a search tree is determined, though we may not know it exactly. Two basic operations help find all potential paths: \textit{\textbf{Forward Search}} and \textit{\textbf{Backward Tracing}}. In \textit{\textbf{Forward Search}}, all subsets of links/edges on the tree are found (Fig. \ref{fig:path_find} (b)); In \textit{\textbf{Backward Tracing}}, the found links/edges are daisy-chained to form all feasible paths (Fig. \ref{fig:path_find} (c) ). In this example, \textit{\textbf{Forward Search}} identifies six tree leaves, six paths are found after performing \textit{\textbf{Backward Tracing}}, and then the path with the highest score (solid lines on Fig. \ref{fig:path_find} (c) ) is the winner for the given strategy. At most $I$ potential paths can be found by performing the above process for all $I$ strategies, and finally the one with highest score along with the strategy is selected.

\textbf{Complexity Analysis}: The computational complexity is graph dependent, and obtaining an exact value for a specific graph is difficult and does not provide much insight. Instead, let us consider a worst-case, i.e., a fully-connected graph $G_1({\cal N}_1, {\cal E}_1)$ with $M_1$ nodes and $M_1(M_1-1)$ edges. Refer to the algorithm, \textit{\textbf{Forward Search}} needs to check edges up to $I\times n_{max}\times M_1(M_1-1)$ times, and \textit{\textbf{Backward Tracing}} needs to do edge connecting for up to $I\times J_{max}\times (n_{max}-1)$ times, where $n_{max}=max(n_1,n_2,\cdots,n_I)$ and $J_{max}=max(J_1,J_2,\cdots,J_I)$. Since $J_i\le M_1(M_1-1),\;i=1,2,\cdots,I$, and $J_{max}\le M_1(M_1-1)$, the upper-bound algorithm complexity is approximately in proportion to $I\times n_{max}\times M_1(M_1-1)\approx I\times n_{max}\times M_1^2$, or in the order of ${\cal O}(I\times n_{max}\times M_1^2)$. This indicates that the graph size ($M_1$) has significant impact on the complexity and suggests that some countermeasures to tackle the exponential increment of complexity would be necessary. A natural idea in the D2D environment is to consider some bounding conditions based on geographical or social relationship in forming a cohort of workers for a given requested computation task. Another idea is graph trimming to be introduced below.

\subsection{Graph Trimming Method}
An exhaustive search eventually finds the best path and its associated configuration strategy, but, as implied above, computational complexity can be a heavy burden to prevent using the path finding algorithm. We can trim the graph globally to cut out unlikely paths, leading to a decreased number of paths in the search space and a reduction on the total computational complexity. This may be done periodically and the obtained result can be shared among all potential requesters.

In addition to link quality and reliability, the computing resource at each participant needs to be considered as well. The resources available at node $k,\;k=1,2,\cdots,|{\cal N}_0|$, can be represented by a resource index vector $\boldsymbol{u}_k=(c_k, m_k, b_k, w_k)$, or a scalar $u_k =\boldsymbol{u}_k\boldsymbol{\rho}^T$ with $\boldsymbol{\rho}=(\rho_1,\rho_2,\rho_3,\rho_4)$ 
being a predefined weighting vector. Similar to link reliability, we introduce a concept of link resource:
\begin{eqnarray}
	&& u_{s,t} = \sqrt{u_s \cdot u_t},\; u_s,u_t\in [0,1]
\end{eqnarray}
To reflect the total effect of link quality, node reliability and node resource, we further introduce a parameter of joint link weight defined as
\begin{eqnarray}
	L_{s,t} = q_{s,t}\cdot r_{s,t}\cdot u_{s,t}
\end{eqnarray}

Let $A=[a_{s,t}]$, $a_{s,t} \in \{0, 1\}$, be a $M_1\times M_1$ adjacency matrix for graph $G_1({\cal N}_1, {\cal E}_1)$ with $M_1=|{\cal N}_1|$ nodes, where $a_{s,t}=1$ means link $(s,t)$ meets link requirement, while $a_{s,t}=0$ means the link is not usable. Practically, with $\theta$, a predefined threshold on the joint link weight, the adjacency matrix $A$ can be expressed as
\begin{eqnarray}
	&& A=[a_{s,t}], \; a_{s,t}=
	\begin{cases}
	0, & \mbox{if } L_{s,t} < \theta \\
	1, & \mbox{if } L_{s,t} \ge \theta
	\end{cases} \\
	 &&\hspace{10mm} s,t=1,2,\cdots,M_1 \nonumber
\end{eqnarray}

Suppose we are forming a length-1 pipeline with just one link (and two nodes), a requester only needs to search its nearby trusted workers that are directly connected to the requester, and the outcome is the best link among all of the direct links, i.e., the one with the highest reliability. To find a length-1 pipeline for all possible requesters, the worker cohort can be simply found by using the adjacency matrix $A=[a_{s,t}]$ directly in a 2-step graph trimming process:
\begin{eqnarray}
	&& \hspace{-6mm} \mbox{Initialization: } {\cal N}=\Phi,\; {\cal E}= {\cal E}_1 ; \nonumber\\
	&& \hspace{-6mm} \mbox{Step 1: For all node pairs } (s,t),\;  s,t=1,2,\cdots,M_1,\mbox{ if} \nonumber\\
	&& \hspace{5mm} a_{s,t} = 1, \mbox{ add nodes } s \mbox{ and } t \mbox{ into } {\cal N}; \nonumber\\
	&& \hspace{-6mm} \mbox{Step 2: Remove from } {\cal E} \mbox{ those edges whose end nodes do } \nonumber\\
	&& \hspace{5mm} \mbox{ not belong to } {\cal N} . \nonumber
\end{eqnarray}
The resultant graph $G({\cal N}, {\cal E})$ is a reduced graph and the node set ${\cal N}$ represents the worker cohort.

\begin{figure}[h]
\vspace{0mm}
\setlength{\abovecaptionskip}{1mm}
\setlength{\belowcaptionskip}{1mm}
    \centering
\begin{minipage}{0.5\textwidth}
    \centering
	\includegraphics[width=0.9\textwidth]{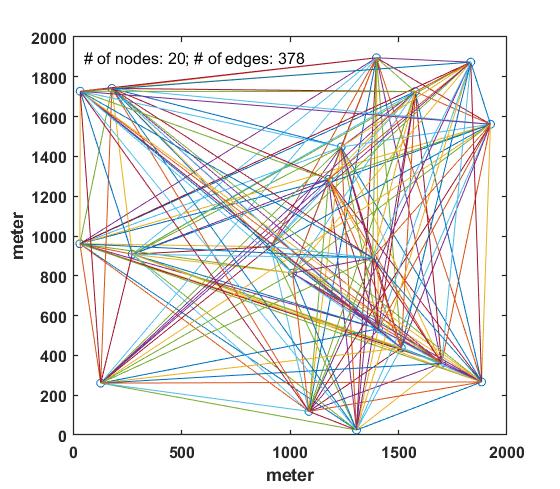}
	\subcaption{Busy graph without trimming.}
\end{minipage}
\begin{minipage}{0.5\textwidth}
    \centering
	\includegraphics[width=0.9\textwidth]{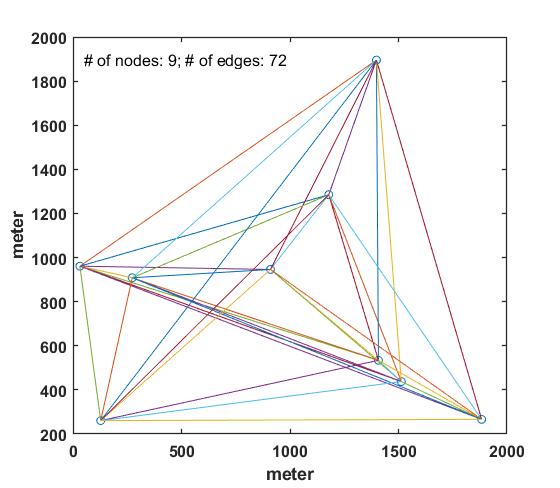}
	\subcaption{Reduced graph obtained by trimming.}
\end{minipage}
	\caption{A graph reduction example.}
\label{fig:graph-reduction}
\end{figure}

However, if a pipeline needs to daisy-chain more than one worker, the adjacency matrix does not give any clue directly for finding a cohort of good candidates. Interestingly, the $n$-th power of the adjacency matrix $A^n$ has some helpful property \cite{selinski2014method,mukherjee2014role}: the $(s,t)$-th entry of $A^n$ gives the number of paths with $n$ connected edges between node $s$ and node $t$, where a counted path may be part of a longer path (i.e., a path with more than $n$ edges). This property inspires a very selective graph trimming approach. Given strategy $i$, our desire is to only consider those paths that are at least $n_i$-edges long, though this desire might be too ideal to implement.
If the value of the $(s,t)$-th entry of $A^n$ is high, then it is more probable that the associated two nodes are on a path with at least $n$ edges. 
Based on this rationale, we propose a generalized graph trimming rule that is similar to the above 2-step process but uses a threshold $\eta$ to filter out all node pairs corresponding to small-value entries in $A^n$.
One remaining issue is selection of parameter $n$ in $A^n$, and we propose a practical way to take into account different powers of $A$ in a weighing fashion. The revised trimming rule follows:
\begin{eqnarray}
	&& \hspace{0mm} \mbox{\textbf{Graph trimming rule ($G_1({\cal N}_1, {\cal E}_1) \longrightarrow G({\cal N}, {\cal E})$):}}\nonumber\\
	&& \hspace{-6mm} \mbox{Initialization: } {\cal N}=\Phi,\; {\cal E}= {\cal E}_1 ; \nonumber\\
	&& H=[h_{s,t}]=\sum_{n=2}^{n_{max}} \alpha_{n-1} A^n, \; n_{max}=max\{n_i\} ; \\
	&& \hspace{-6mm} \mbox{Step 1: For all node pairs } (s,t),\;  s,t=1,2,\cdots,M_1, \mbox{ if} \nonumber\\
	&& \hspace{5mm} h_{s,t} \ge \eta, \mbox{ add nodes } s \mbox{ and } t \mbox{ into } {\cal N}; \nonumber\\
	&& \hspace{-6mm} \mbox{Step 2: Remove from } {\cal E} \mbox{ those edges whose end nodes do } \nonumber\\
	&& \hspace{5mm} \mbox{ not belong to } {\cal N} . \nonumber
\end{eqnarray}
where $H=[h_{s,t}]$ is a $M_1\times M_1$ matrix with entries $h_{s,t}, \;s,t=1,2,\cdots,M_1$, $\alpha_n \ge 0$ are customized weights, and $\eta$ is a predefined threshold. The effectiveness of the proposed trimming method can be visualized with an example shown in Fig. \ref{fig:graph-reduction}.

\section{Preliminary Assessment of Feasibility}
Laboratory experiments and computer simulations have been used to examine the feasibility of the proposed concept, which prepares us for prototyping a sophisticated mobile pipeline computing system in the near future. We built a small testbed to mimic a pipeline of computing engines connected via D2D communication, whereas all D2D links between devices are emulated by using WiFi protocol.

\vspace{0mm}
\subsection{Experiment Setup}
\vspace{0mm}




\begin{figure}
\vspace{3mm}
\setlength{\abovecaptionskip}{2mm}
\setlength{\belowcaptionskip}{2mm}
    \centering
\begin{minipage}{0.5\textwidth}
    \centering
	\includegraphics[width=0.95\textwidth]{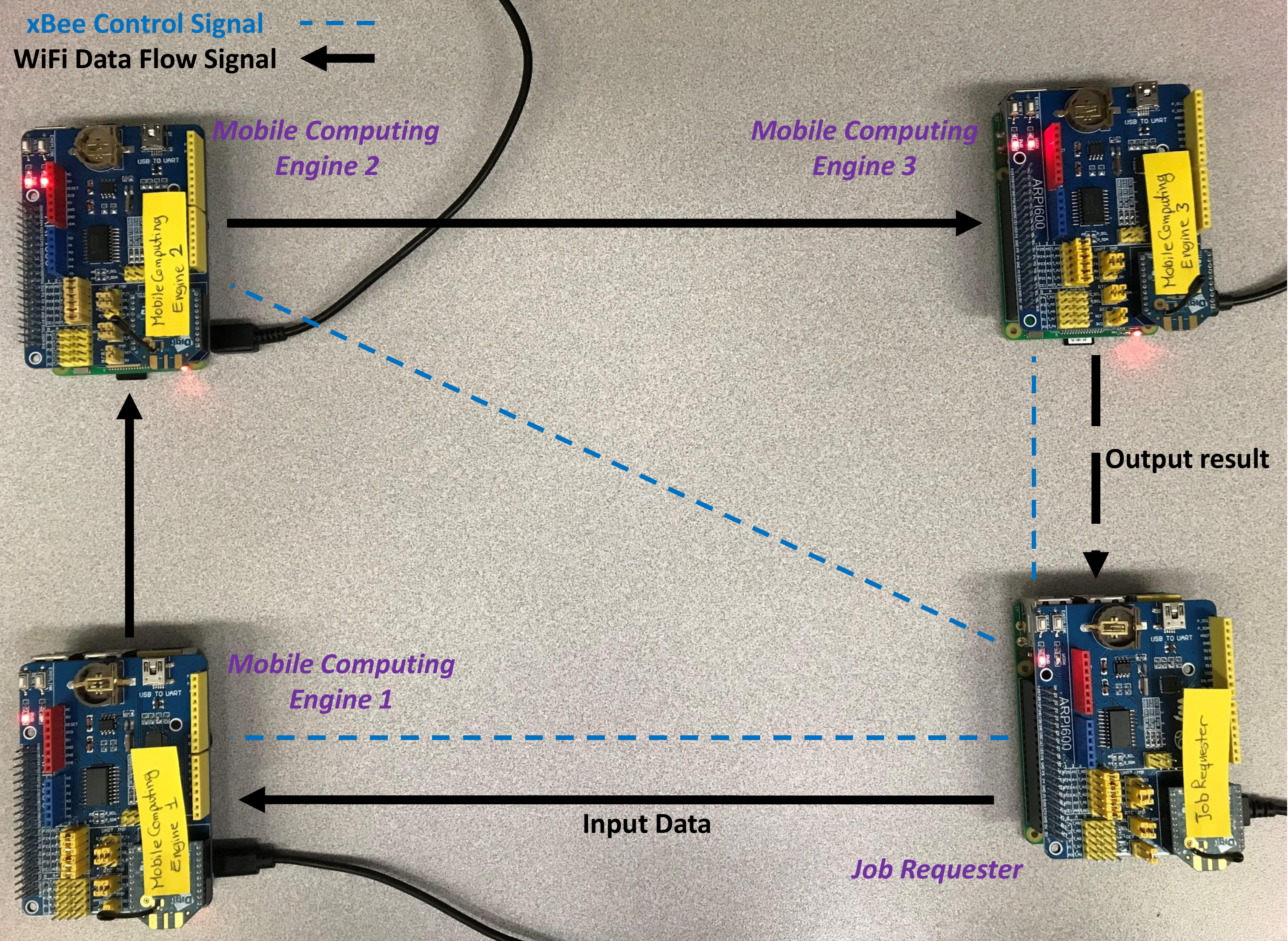}
	\subcaption{A 3-worker pipeline computing setup using WiFi and ZigBee communication protocols.}
\end{minipage}
\begin{minipage}{0.5\textwidth}
\vspace{3mm}
    \centering
	\includegraphics[width=0.65\textwidth,trim=0 0 0 4,clip]{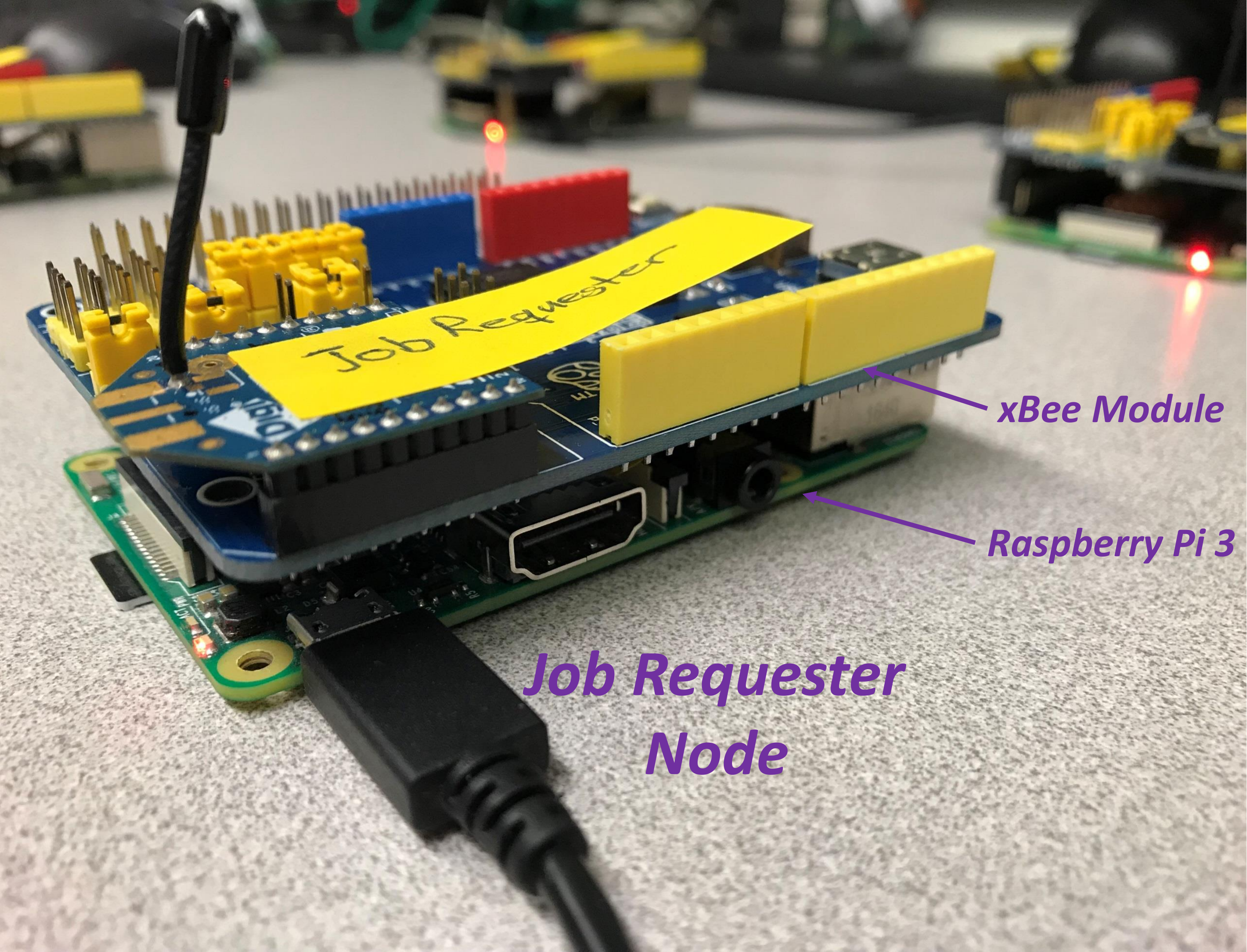}
	\subcaption{Each of workers and requester is a combination of Raspberry Pi 3 and XBee modules.}
\end{minipage}
	\caption{Experiment setup.}
\label{fig:setup}
\end{figure}

Shown in Fig. \ref{fig:setup} is a setup mimicking a real-world scenario of mobile pipeline computing, where each Raspberry Pi 3 module integrated with a XBee module is used to represent either a requester or a worker. The wireless LAN interface (WiFi) on the raspberry Pi 3 devices has options IEEE 802.11 b/g/n, and 802.11n is used in this experiment to emulate D2D  for transmitting wideband processed data. On the other hand, ZigBee\footnote{ZigBee is a Home Area Network (HAN) protocol built upon the 802.15. 4 IEEE standard, whereas XBee is a brand name referring to a family of devices from Digi International that support a group of HAN protocols including ZigBee and its enhanced versions.} communication protocol is dedicated for transmitting narrowband control messages. Each of worker and requester devices is linked to its previous and next devices using WiFi protocol, and the requester is connected to all the workers using ZigBee protocol.

\begin{figure*}[t]
\vspace{0mm}
\setlength{\abovecaptionskip}{0mm}
\setlength{\belowcaptionskip}{3mm}
    \centering
\begin{minipage}{1\textwidth}
    \centering
	\includegraphics[width=1\textwidth]{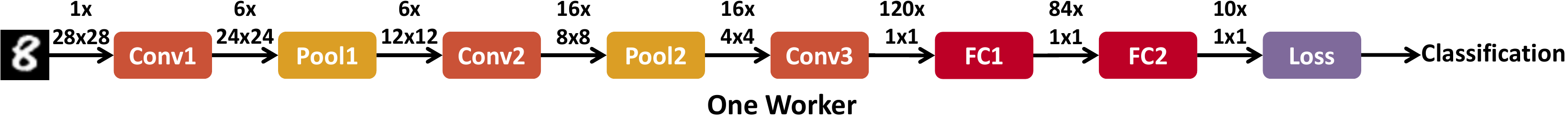}
	\subcaption{Case I, all LeNet neural network layers are run on one worker.}
	\label{fig:OneDevice}
\end{minipage}
\begin{minipage}{1\textwidth}
    \centering
	\includegraphics[width=1\textwidth]{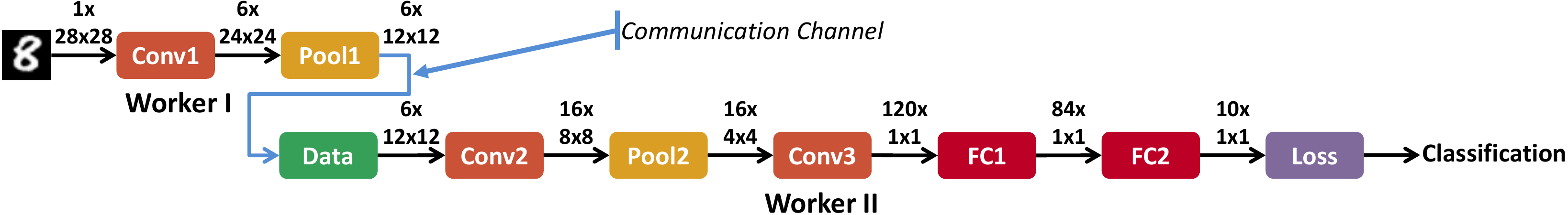}
	\subcaption{Case II, LeNet layers are divided between and run on two workers.}
	\label{fig:TwoDevices}
\end{minipage}
\begin{minipage}{1\textwidth}
    \centering
	\includegraphics[width=1\textwidth]{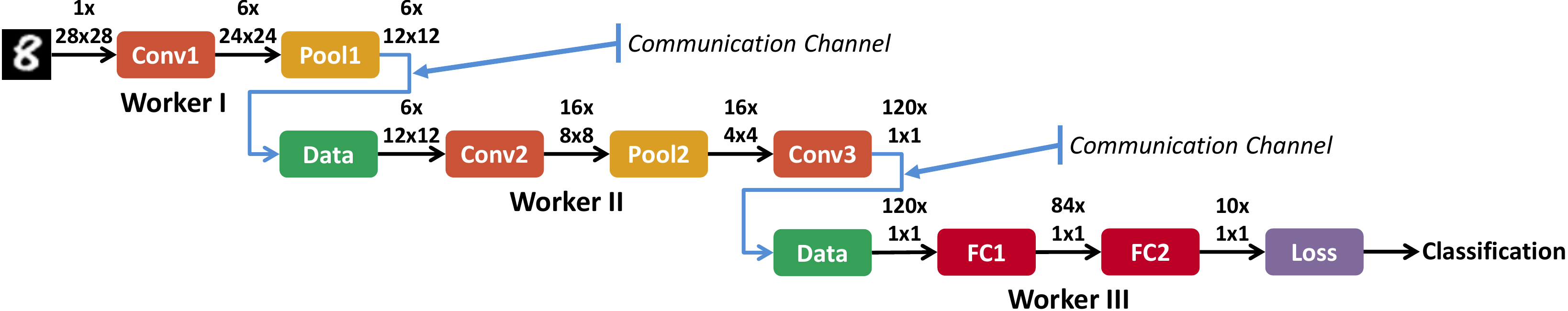}
	\subcaption{Case III, LeNet layers are divided among and run on three workers.}
	\label{fig:ThreeDevices}
\end{minipage}
	\caption{LeNet neural network layer partitioning with different numbers of workers. Conv = Convolution layer; Pool = Pooling layer; FC = Fully Connected layer; Loss = Loss Function.}
	\vspace{0mm}
\label{fig:pipeline}
\end{figure*}

%
%

\vspace{0mm}
\subsection{Demonstration of Pipeline Computing}
\label{Effectiveness of The Pipeline Computing}
\vspace{0mm}

It is expected that by taking advantage of available spare computing resources, overall run time can be reduced. However this needs to be validated experimentally since an accurate outcome affected by various practical factors cannot be quantified theoretically. A small-scale deep learning task is tested on this setup by considering two types of operations: non-pipeline and pipeline operation making use of additional computing resources. In general, a computational process needs to be partitioned into subprocesses, and these subprocesses are sequentially arranged and assigned to distributed workers connected in a daisy-chain manner. Luckily, in Deep Neural Network (DNN) case, a computational process can be naturally divided into sequential subprocesses since a DNN structure is actually formed in a sequential manner (see Fig. \ref{fig:pipeline}). 
In non-pipeline operation (Case I in Fig. \ref{fig:OneDevice}), there is only one worker node on which the whole LeNet DNN \cite{lecun1998gradient} is implemented to classify the images, as shown in Fig. \ref{fig:OneDevice}.
In pipeline operation (Case II and Case III in Figs. \ref{fig:TwoDevices} and \ref{fig:ThreeDevices}, respectively), the LeNet DNN is distributed among the workers to classify the images in a coordinated way to increase process throughput.

\begin{table}[t]
  \centering
  \caption{Measured result for use of different number of workers.}
    \begin{tabular}{|c|c|c|c|c|}
    \hline
    \hspace{0mm} Case \hspace{0mm} & \hspace{0mm} \# of workers \hspace{0mm} & \hspace{0mm} Total time (ms) \hspace{0mm} & \hspace{0mm} Throughput \hspace{0mm} \\
    \hline
    I & One & 540.103 & 100\% \\
    \hline
    II & Two & 347.780 & 155\% \\
    \hline
    III & Three & 308.457 & 175\% \\
    \hline
    \end{tabular}%
    \vspace{0mm}
  \label{tab:throughput}%
\end{table}%

Shown in Table \ref{tab:throughput} is the testing result. When the LeNet DNN ran on one device (non-pipeline, Case I), it took a total time of about 540 ms to classify 100 images, where the total time includes processing time and communication time.
To classify the same images using pipeline operation, with two and three workers it took about 348 ms and 308 ms, respectively. In other words, pipeline based LeNet DNN can increase classification throughput from 100\% to 155\% for two workers and to 175\% for three workers.
The experiment result agrees with our expectation that the more workers are involved the better performance is achieved.
This experiment implies that pipeline computing on coordinated devices in a D2D-enable environment is feasible. Although this experiment has a limited scope, one can envision the same trend in performance improvement for different computation levels.

\vspace{-3mm}
\begin{table}[t]
\caption{Strategy table made from experiment.}
\begin{center}
\begin{tabular}{|c|c|c|c|c|c|}
\hline
	{\tiny\bf Strategy} & {\tiny\bf \# of} & {\tiny\bf Computing} & {\tiny\bf Memory} & {\tiny\bf Buffer} & {\tiny\bf Bandwidth}\\
	{\tiny\bf score} & {\tiny\bf workers} & {\tiny\bf (MAC/s)} & {\tiny\bf (kB)} & {\tiny\bf (kB)} & {\tiny\bf (kB/s)} \\
\hline
	\multirow{3}{*}{0.0989} & \multirow{3}{*}{2} & 0.0 & 0.0 & 0.0 & 1,834.0 \\
\cline{3-6}
	&  & 10,164.7	& 15.6	& 313.6	& 4,361.0 \\
\cline{3-6}
	&  & 15,363.1 &	4,427.0	& 1,382.4	& 60.0 \\
\hline
	\multirow{3}{*}{0.2248} & \multirow{3}{*}{2} & 0.0 & 0.0 & 0.0 & 1,834.0 \\
\cline{3-6}
	&  & 10,096.6 & 15.6 & 313.6 & 1,623.0 \\
\cline{3-6}
	&  & 18,172.2 & 4,427.0 & 345.6 & 60.0 \\
\hline
	\multirow{3}{*}{0.2511} & \multirow{3}{*}{2} & 0.0 & 0.0 & 0.0 & 1,834.0 \\
\cline{3-6}
	&  & 18,872.9 & 257.6 & 313.6 & 1,721.0 \\
\cline{3-6}
	&  & 5,925.0 & 4,185.0 & 409.6 & 60.0 \\
\hline
	\multirow{3}{*}{0.3204} & \multirow{3}{*}{2} & 0.0 & 0.0 & 0.0 & 1,834.0 \\
\cline{3-6}
	&  & 18,521.2 & 257.6 & 313.6 & 620.0 \\
\cline{3-6}
	&  & 6,712.1 & 4,185.0 & 102.4 & 60.0 \\
\hline
	\multirow{3}{*}{0.4107} & \multirow{3}{*}{2} & 0.0 & 0.0 & 0.0 & 1,834.0 \\
\cline{3-6}
	&  & 16,188.2 & 3,341.6 & 313.6 & 471.0 \\
\cline{3-6}
	&  & 3,522.6 & 1,101.0 & 48.0 & 60.0 \\
\hline
	\multirow{3}{*}{0.4681} & \multirow{3}{*}{2} & 0.0 & 0.0 & 0.0 & 1,834.0 \\
\cline{3-6}
	&  & 15,504.4 & 4,357.6 & 313.6 & 494.0 \\
\cline{3-6}
	&  & 466.7 & 85.0 & 33.6 & 60.0 \\
\hline
	\multirow{4}{*}{0.0759} & \multirow{4}{*}{3} & 0.0 & 0.0 & 0.0 & 1,834.0 \\
\cline{3-6}
	&  & 10,096.6 & 15.6 & 313.6 & 1,472.0 \\
\cline{3-6}
	&  & 30,317.7 & 242.0 & 345.6 & 621.0 \\
\cline{3-6}
	&  & 6,716.1 & 4,185.0 & 102.4 & 60.0 \\
\hline
	\multirow{4}{*}{0.1435} & \multirow{4}{*}{3} & 0.0 & 0.0 & 0.0 & 1,834.0 \\
\cline{3-6}
	&  & 10,096.6 & 15.6 & 313.6 & 1,472.0 \\
\cline{3-6}
	&  & 22,602.4 & 3,326.0 & 345.6 & 471.0 \\
\cline{3-6}
	&  & 3,522.6 & 1,101.0 & 48.0 & 60,0 \\
\hline
\end{tabular}
\end{center}
\label{tab:rawdata}
* MAC/s = Multiply-Accumulate/sec.
\end{table}

\begin{table}[h]
\caption{Some of simulation parameters}
\centering
\begin{tabular}{ccc}
\hline
Parameter & Description & Value \\
\hline
	$\boldsymbol{\rho}$ & resource-weighting vector & (0.4, 0.25, 0.25, 0.1) \\
\hline
	\multirow{2}{*}{$(\alpha_1,\alpha_2,\alpha_3)$} & weights on different powers & \multirow{2}{*}{(0.3, 0.7, 0)} \\
	& of the adjacent matrix & \\
\hline
	\multirow{2}{*}{$(\beta_1,\beta_2,\beta_3)$} & weights on the three & \multirow{2}{*}{(0.05, 0.5, 0.3)} \\
	 & types of scores & \\
\hline
\label{tab:simulation_parameters}
\end{tabular}
\end{table}

\vspace{0mm}
\subsection{Computer Simulation}
\label{Computer Simulation}
\begin{figure}[h]
\vspace{0mm}
	\centering
	\includegraphics[width=0.45\textwidth]{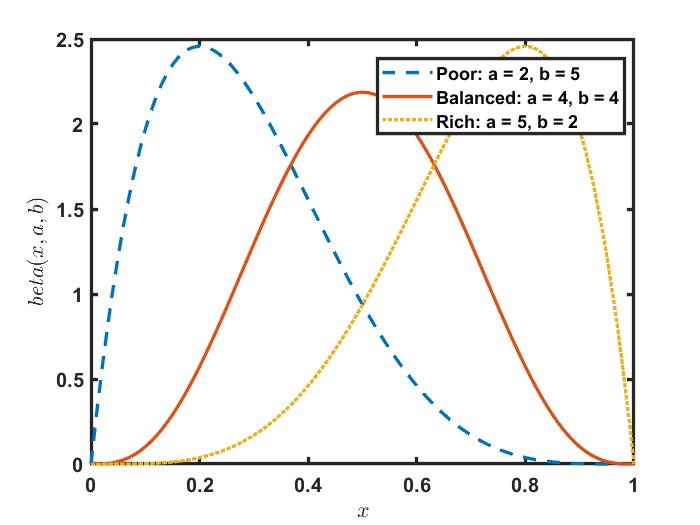}
	\caption{Beta distribution.}
\label{fig:beta}
\end{figure}

\vspace{0mm}
The effectiveness of our proposed path finding techniques is tested using computer simulation. A large number of parameters need to be specified for the simulation. Eight pipeline configuration strategies for a deep learning inference job are provided in Table \ref{tab:rawdata}, where the preference scores are specified based on required resource consumptions, and the required resource consumptions is normalized before feeding into the algorithm. Three types of random variables (RVs) need to be considered: 1) link qualities, 2) node reliabilities, and 3) available resources. An exception is that the requester always sets its reliability level to the highest, i.e., one. For convenience, beta distribution $beta(x, a, b)$ is used to model the probability density functions (pdfs) of these RVs. A modeled pdf can be adjusted flexibly by tuning the two parameters $a$ and $b$. For instance, as shown in Fig. \ref{fig:beta}, three probability distributions are mimicked using three different parameter pairs.

Essential performance metrics include (i) path score $S_P=\max_{i, j\in{\cal P}_i} \left\{\beta_1 Q_j + \beta_2 R_j + \beta_3 f_i \right\}$, (ii) $P_P(1)$, probability that at least one qualified path exists,
and (iii) $R_E$, edge reduction rate contributed by graph trimming. $S_P$, $P_P(1)$ and $R_E$ are affected by pool size ($M$), link quality, node reliability, resource availability and required minimum resources, etc. Table \ref{tab:simulation_parameters} shows some key parameters used in the simulation, where the values of $\beta_1,\beta_2$ and $\beta_3$ are taken such that $\beta_1 Q_p$, $\beta_2 R_p$ and $\beta_3 f_i $ are at comparable levels. How to select these parameters to accurately mimic the reality is beyond the scope of this paper. 
Instead, we consider two system settings for qualitative comparison
: case 1 with relatively small pool size, poor condition and less resources, and case 2 with relatively large pool size, good condition and rich resources. 

\begin{figure}[h]
\vspace{0mm}
\setlength{\abovecaptionskip}{3mm}
\setlength{\belowcaptionskip}{1mm}
	\centering
	\includegraphics[width=0.45\textwidth,trim=4 4 4 4,clip]{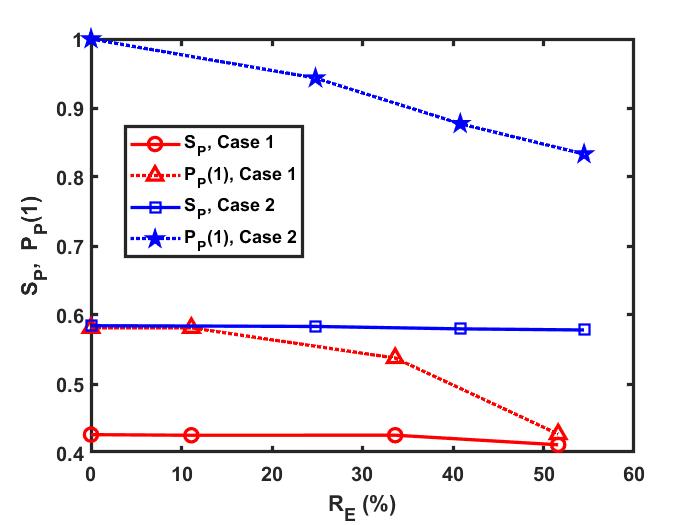}
	\caption{Test result for Case 1: $M = 20, beta(x, 4, 4)$ and Case 2: $M = 28, beta(x, 5, 2)$.}
\label{fig:CaseStudy}
\end{figure}

The winning strategy is strategy 6 which has a higher preference score, and the simulation results for the two cases are provided in Fig. \ref{fig:CaseStudy}. Obviously, Case-1 setting (poor condition) corresponds to a lower score ($S_P$) and worse chance of pipeline existence ($P_P(1)$). One can see that graph trimming (decrease of $R_E$) dose not cause noticeable degradation in the optimality ($S_P$); and if $R_E$ is kept below 20\%, the reduction in the chance of pipeline existence ($S_P(1)$) is negligible. Note that a shorter pipeline is more likely selected than a longer one, since a shorter path is in favor of path quality and reliability; but the likelihood can be adjusted via assigning a higher strategy score ($f_{i'}$) to a preferred strategy ($i'$). Furthermore, the amount of edge reduction can be tuned by changing the thresholds $\theta$ and/or $\eta$.
\begin{figure}[h]
\vspace{3mm}
\setlength{\abovecaptionskip}{3mm}
\setlength{\belowcaptionskip}{0mm}
\centering
    \includegraphics[width=0.5\textwidth,trim=4 4 4 4,clip]{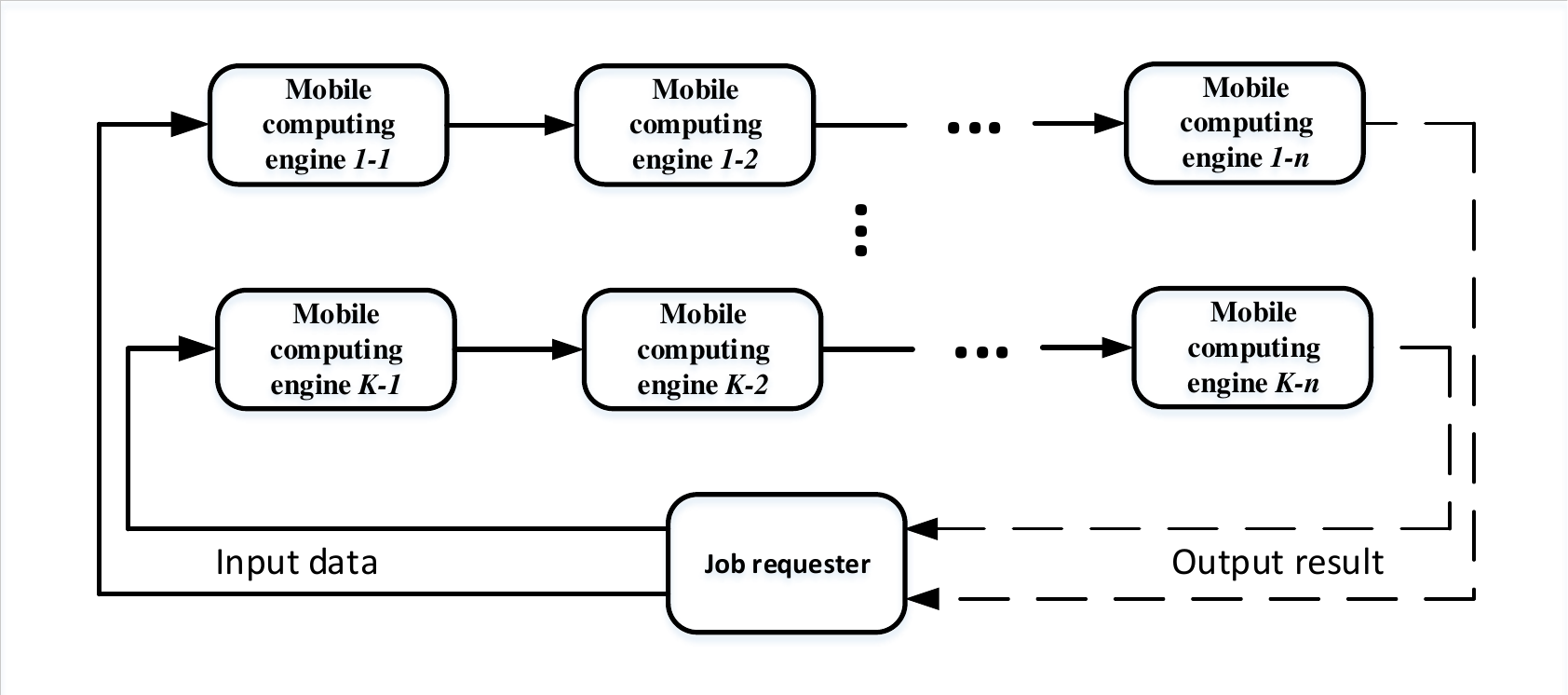}
    \caption{$K$-concurrent pipeline computing concept.}
\label{fig:ConcurrentPipeline}
\end{figure}

\section{Stability Analysis of mmWave D2D Based Opportunistic Mobile Pipeline Computing}
5G technology integrated with mmWave enables GHz-bandwidth transmission at millisecond-level latency. Since D2D is part of 5G, in the near future our proposed mobile pipeline computing can enjoy the advantage of large bandwidth and low-latency offered by 5G mmWave. However, mmWave communication is very sensitive to link blockage \cite{collonge2004influence,singh2009blockage,jung2016connectivity,maccartney2017rapid}, and the blockage effect will be amplified as multiple links have to be utilized simultaneously in the pipeline computing scenario. In mmWave communication, blockage is a dominating factor responsible for interruption of a communication session. Therefore, in performance analysis it is reasonable to 
ignore other possible factors that may make contribution to the interruption of a pipeline session. In this section we analyze the pipeline stability using a 2-state discrete-time Markov chain as a dynamic blockage model, assuming all links behave statistically identically and independently. In \cite{maccartney2017rapid} the 2-state blockage model is validated based on measurement. The model includes two states called ``Unshadowed'' (or ${\cal G}$ standing for Good) and ``Shadowed'' (or ${\cal B}$ standing for Bad), along with a $2\times 2$ transition probability matrix. From a pipeline stability perspective, we are concerned about the following two parameters given that a pipeline has been formed:
\begin{enumerate}
	\item The probability of success (i.e., a computation job is not interrupted by blockage) for a given session time;
	\item The average number of attempts to form a pipeline for a requested computation job, assuming the statistic model keeps unchanged and the requester continues to request a pipeline computing session till the job is completed.
\end{enumerate}

Let $T$ be the session time defined as a duration that starts upon a pipeline is formed and ends right after the computation job is completed, $n_{node}$ the required number of nodes (including both requester and workers) on a pipeline, $P_S(T,1)$ the probability of success conditioned on that a pipeline is formed, $\Delta t$ ($\ll T$) the sampling time interval of the dynamic blockage model, and $\epsilon$ the probability for transition from state ${\cal G}$ to state ${\cal B}$, i.e., $\epsilon=Pr({\cal B}|{\cal G})$. Using the (1st-order) Markov chain property and denoting $m=round(T/\Delta t)$, $P_S(T)$ is given by
\begin{eqnarray}
	P_S(T,1) &\hspace{-3mm}=&\hspace{-3mm} \left[Pr({\cal G}^m|{\cal G})\right]^{n_{node}-1} \nonumber\\
	&\hspace{-3mm}=&\hspace{-3mm} \left[Pr({\cal G}^{m+1})/Pr({\cal G})\right]^{n_{node}-1} \nonumber\\
	&\hspace{-3mm}=&\hspace{-3mm} \left[Pr({\cal G}|{\cal G}^m)Pr({\cal G}^m)/Pr({\cal G})\right]^{n_{node}-1} \nonumber\\
	&\hspace{-3mm}=&\hspace{-3mm} \left[Pr({\cal G}|{\cal G})Pr({\cal G}^m)/Pr({\cal G})\right]^{n_{node}-1} \nonumber\\
	&\hspace{-3mm}=&\hspace{-3mm} \left[Pr({\cal G}|{\cal G})Pr({\cal G}|{\cal G}^{m-1})Pr({\cal G}^{m-1})/Pr({\cal G})\right]^{n_{node}-1} \nonumber\\
	&\hspace{-3mm}=&\hspace{-3mm}  \cdots \nonumber\\
	&\hspace{-3mm}=&\hspace{-3mm} \left[Pr({\cal G}|{\cal G})\right]^{m (n_{node}-1)} \nonumber\\
	&\hspace{-3mm}=&\hspace{-3mm} (1-\epsilon)^{m (n_{node}-1)} \nonumber\\
	&\hspace{-3mm}\approx&\hspace{-3mm} (1-\epsilon)^{T (n_{node}-1)/\Delta t}
\end{eqnarray}
where ${\cal G}^m$ stands for ${\cal G}{\cal G}\cdots{\cal G}$, or ``event ${\cal G}$ occurs consecutively for $m$ times'', and $Pr({\cal G}|{\cal G})=1-Pr({\cal B}|{\cal G})=1-\epsilon$ has been applied.

To increase the success probability, a straightforward strategy is to employ multiple pipelines concurrently (Fig. \ref{fig:ConcurrentPipeline}) at increased resource consumption, if these pipelines are available. When $K$ concurrently qualified pipelines are used to execute a job at the same time, assuming the pipelines are statistically independent, the success probability conditioned on that at least $K$ qualified pipelines are formed, denoted by $P_S(T, K)$, is given by
\begin{eqnarray}
	P_S(T, K) &\hspace{-3mm}=&\hspace{-3mm} 1 - (1 - P_S(T,1))^K
\end{eqnarray}

With $P_S(T, K), \; K\ge 1$, it is not difficult to derive the conditional average number of attempts denoted by $\bar{n}(T, K)$,
\begin{eqnarray}
	\bar{n}(T, K) &\hspace{-3mm}=&\hspace{-3mm} \sum_{l=1}^{\infty} l [1-P_S(T, K)]^{l-1} P_S(T, K)\nonumber\\
	&\hspace{-3mm}=&\hspace{-3mm} 1/P_S(T, K)
\end{eqnarray}

\begin{figure}[t]
\vspace{0mm}
\setlength{\abovecaptionskip}{2mm}
\setlength{\belowcaptionskip}{0mm}
\centering
\includegraphics[width=0.45\textwidth]{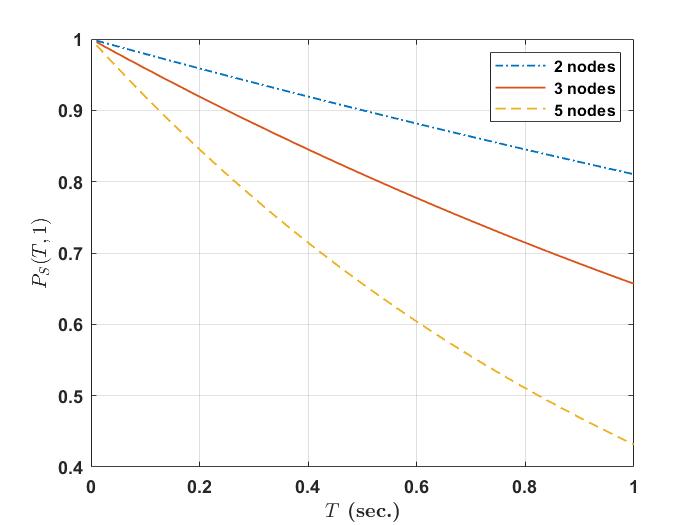}
\caption{Conditional probability of success vs. session time for single pipeline.}
\label{fig:P_S(T1)}
\end{figure}

\begin{figure}[H]
\vspace{0mm}
\setlength{\abovecaptionskip}{2mm}
\setlength{\belowcaptionskip}{0mm}
\centering
\includegraphics[width=0.45\textwidth]{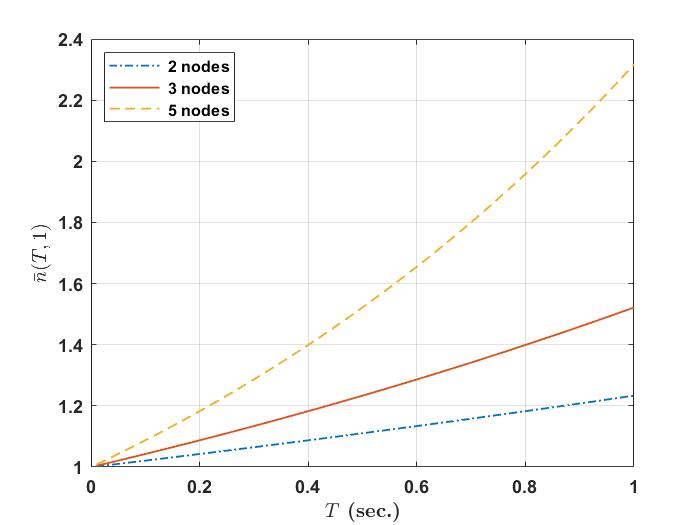}
\caption{Conditional average number of attempts vs. session time for single pipeline.}
\label{fig:n_avg(T1)}
\end{figure}

\begin{figure}[h]
\vspace{0mm}
\setlength{\abovecaptionskip}{2mm}
\setlength{\belowcaptionskip}{0mm}
	\centering
	\includegraphics[width=0.45\textwidth]{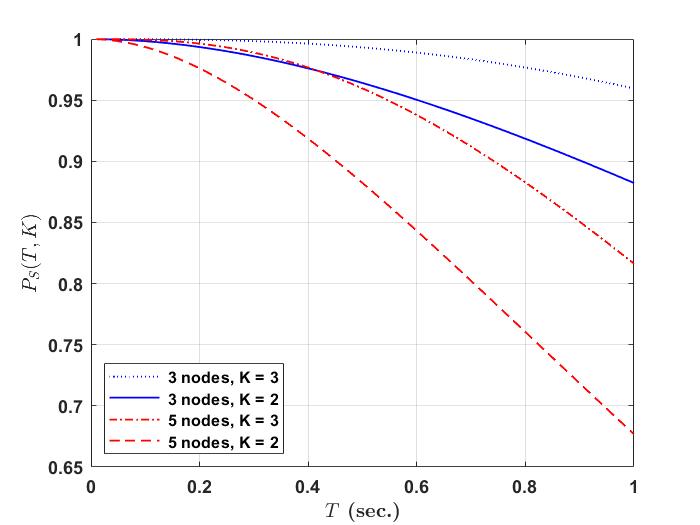}
	\caption{Conditional probability of success vs. session time for concurrent multiple pipelines.}
\label{fig:P_S(TK)}
\end{figure}

\begin{figure}[h]
\vspace{0mm}
\setlength{\abovecaptionskip}{2mm}
\setlength{\belowcaptionskip}{0mm}
	\centering
	\includegraphics[width=0.45\textwidth]{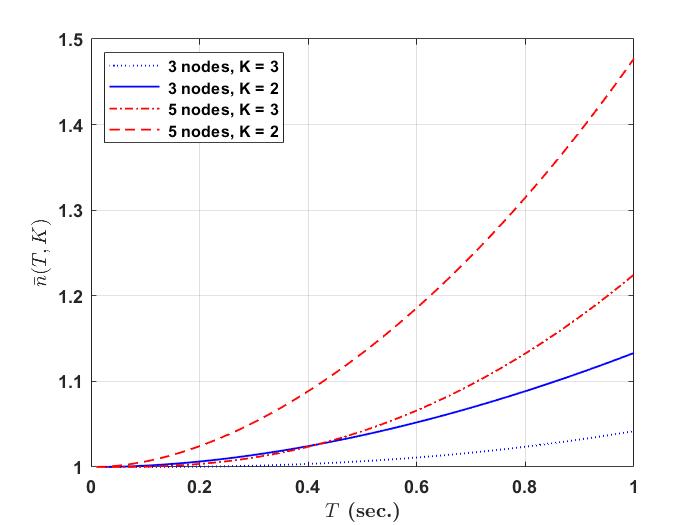}
	\caption{Conditional average number of attempts vs. session time for concurrent multiple pipelines.}
\label{fig:n_avg(TK)}
\end{figure}

Shown in Fig. \ref{fig:P_S(T1)} -- Fig. \ref{fig:n_avg(TK)} are evaluation results using some values provided in \cite{maccartney2017rapid}: $\Delta t$ (being $T$ in \cite{maccartney2017rapid}$) = 3.3$ ms, $\epsilon$ (being $p$ in \cite{maccartney2017rapid}$) = 6.93\times10^{-4}$. As expected, a shorter session time and/or a shorter pipeline lead to a better stability performance, and the use of concurrent multiple pipelines can improve the performance significantly at the cost of increased system resource consumption. Note that, to evaluate stability, all of these results are conditioned on that the pipelines are formed. Although $P_P(K)$, the prior probability that at least $K$ pipeline paths exist depends on multiple factors, it can be expected that this probability would approach one as the size of a mobile computing pool increases.

\section{Conclusions}
The mobile collaborative pipeline computing concept is proposed and studied. Our analytical and experimental results give us confidence on feasibly applying our proposed techniques to real world problems, such as deep learning inference on mobile devices in a D2D environment. Our proposed system should be suitable for real-time on-site computation-intensive tasks for which current cloud computing technology may not be suitable. Based on the foundation laid, further research and development are expected. Future work includes 1) extending single-request single-session service to multi-request multi-session service, considering uncertain resource availability \cite{javadi2011discovering,lazaro2012long,javadi2013modeling}, which resorts to optimum scheduling and system resource management to maximize the overall performance of a community computing pool; 2) improve the path finding framework by taking into account the impact of session time $T$ to maximize the success probability $P_S(T, K)\cdot P_P(K)$ (and minimize $\bar{n}(T, K)\cdot P_P(K)$ at the same time);  3) performance evaluation considering realistic system setting and protocols; and 4) broad issues around computing pool management, such as worker trustworthiness, privacy of computation jobs and the job owners (requesters), and decentralized management.

\bibliography{mybib}

\begin{IEEEbiography}[{\includegraphics[width=1in,height=1.3320in]{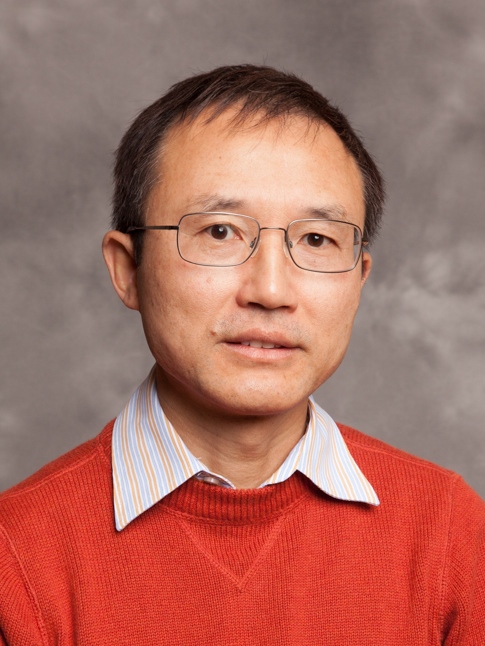}}]{Terry N. Guo} received his M.S. degree in telecommunications engineering from Beijing University of Posts and Telecommunications, Beijing, in 1990, and Ph.D. degree in communications and electronic systems from the University of Electronic Science and Technology of China, Chengdu, in 1997. From January 1997 to December 1999, he was a postdoctoral researcher at the Center for Wireless Communications, University of California, San Diego. He worked for a few startups in New Jersey in early 2000s. Since 2004, he has been with the Center for Manufacturing Research, Tennessee Technological University, Tennessee, playing a wide range of duties including research, teaching and laboratory management. He has been conducting research and prototyping research testbeds in the areas of wireless communications, Radio Frequency (RF) systems, wide band beamforming, statistic signal processing, and data analytics. His recent research interests include Internet of Things (IoT) security and privacy, smart manufacturing, 5G Device-to-Device (D2D) communications, and mobile opportunistic computing.
\end{IEEEbiography}

\begin{IEEEbiography}[{\includegraphics[width=1in,height=1.25in,clip,keepaspectratio]{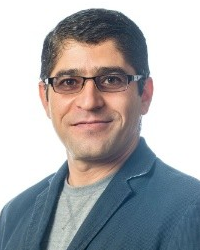}}]
{Hawzhin Mohammed} received his B.Sc. degree in electrical engineering from Salahaddin University, Erbil, Iraq, in 2000. He received his M.Sc. degree from Tennessee Technological University, Cookeville, TN, USA, in 2017, where he is currently pursuing his Ph.D. degree at the Department of Electrical and Computer Engineering. His current research interest includes wireless network security, hardware security, IoT security, machine learning, and deep learning.
\end{IEEEbiography}

\begin{IEEEbiography}[{\includegraphics[width=1in,height=1.25in,clip,keepaspectratio]{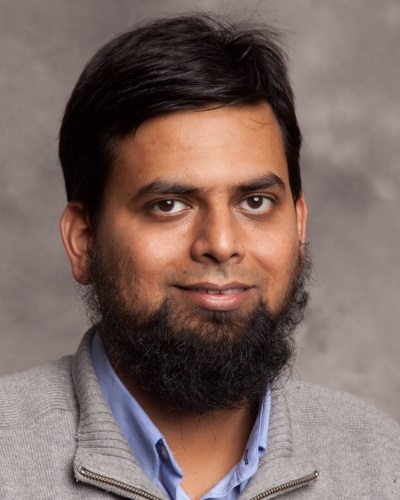}}]
{Syed Rafay Hasan} received the B.Eng. degree in electrical engineering from the NED University of Engineering and Technology, Pakistan, and the M.Eng. and Ph.D. degrees in electrical engineering from Concordia University, Montreal, QC, Canada. From 2006 to 2009, he was an Adjunct Faculty Member with Concordia University. From 2009 to 2011, he was a Research Associate with the Ecole Polytechnique de Montreal. Since 2011, he has been with the Electrical and Computer Engineering Department, Tennessee Tech University, Cookeville, TN, USA, where he is currently an Associate Professor. He has published more than 69 peer-reviewed journal and conference papers. His current research interests include hardware design security in the Internet of Things (IoT), hardware implementation of deep learning, deployment of convolution neural networks in the IoT edge devices, and hardware security issues due to adversarial learning. He received the Postdoctoral Fellowship Award from the Scholarship Regroupment Stratgique en Microsystmes du Québec, SigmaXi Outstanding Research Award, Faculty Research Award from Tennessee Tech University, the Kinslow Outstanding Research Paper Award from the College of Engineering, Tennessee Tech University, and the Summer Faculty Fellowship Award from the Air force Research Lab (AFRL). He has received research and teaching funding from NSF, ICT-funds UAE, AFRL, and Intel Inc. He has been part of the funded research projects, as a PI or a Co-PI, that worth more than \$1.1 million. He has been the Session Chair and Technical Program Committee Member of several IEEE conferences including ISCAS, ICCD, MWSCAS, and NEWCAS, and a Regular Reviewer for several IEEE Transactions and other journals including TCAS-II, IEEE ACCESS, Integration, the VLSI Journal, IET Circuit Devices and Systems, and IEEE Embedded System Letters.
\end{IEEEbiography}

\end{document}